\documentclass[a4paper,superscriptaddress,preprint,groupedaddress]{revtex4-1}
\usepackage{graphicx} 
\usepackage{hyperref}
\usepackage{caption}
\usepackage{color}
\usepackage{stackengine}
\usepackage[position=t,singlelinecheck=off]{subcaption}
\usepackage{array}
\usepackage{multirow}
\usepackage{enumerate}
\usepackage{amsmath}
\usepackage{amssymb}
\usepackage{amsthm}
\usepackage{amsfonts}
\usepackage{ascmac}
\usepackage[nameinlink]{cleveref} 
\usepackage{url}
\usepackage{bm}
\usepackage[utf8]{inputenc}
\usepackage[T1]{fontenc}
\usepackage[scaled]{helvet}%
\usepackage{lmodern}
\crefname{equation}{Eq.\!}{Eqs.\!}
\crefname{figure}{Fig.\!}{Figs.\!}
\usepackage{blindtext}
\usepackage{microtype} 
\usepackage{lineno} 
\makeatletter
\newcommand*{\rom}[1]{\expandafter\@slowromancap\romannumeral #1@}
\makeatother
\def\conf{{\hspace{0.1em}\text{conf}\hspace{0.1em}}}

\usepackage{xcolor}
\hypersetup{
  colorlinks   = true, 
  urlcolor     = black, 
  linkcolor    = blue, 
  citecolor   = blue 
}
\def\Qcp{Q^{\text{cp}} _{\text{config}}}
\def\ER{Erd\H{o}s-R\'{e}nyi\ }
\def\Exp{{\mathbb E}}
\def\Amat{\bm A}
\def\Ymat{\bm Y}

\newcommand\given[1][]{\:#1\vert\:}
\definecolor{purple}{RGB}{170, 0, 255}
\begin{document}
\title{Core-periphery structure requires something else in the network}
\author{Sadamori Kojaku}
\author{Naoki Masuda}
\email{naoki.masuda@bristol.ac.uk}
\affiliation{
Department of Engineering Mathematics,
Merchant Venturers Building, University of Bristol,
Woodland Road, Clifton, Bristol BS8 1UB, United Kingdom
}
\begin{abstract}
A network with core-periphery structure consists of core nodes that are densely interconnected. 
In contrast to community structure, which is a different meso-scale structure of networks, core nodes can be connected to peripheral nodes and peripheral nodes are not densely interconnected.
Although core-periphery structure sounds reasonable, we argue that it is merely accounted for by heterogeneous degree distributions, if one partitions a network into a single core block and a single periphery block, which the famous Borgatti-Everett algorithm and many succeeding algorithms assume.
In other words, there is a strong tendency that high-degree and low-degree nodes are judged to be core and peripheral nodes, respectively. 
To discuss core-periphery structure beyond the expectation of the node's degree (as described by the configuration model), we propose that one needs to assume at least one block of nodes apart from the focal core-periphery structure, such as a different core-periphery pair, community or nodes not belonging to any meso-scale structure. 
We propose a scalable algorithm to detect pairs of core and periphery in networks, controlling for the effect of the node's degree. 
We illustrate our algorithm using various empirical networks.
\end{abstract}
\maketitle

\section{Introduction}
Many complex systems, biological, physical or social, can be represented by networks \cite{Newman2010,Barabasi2016}. 
A network consists of a set of nodes and edges, where nodes represent objects (e.g., people, web pages) and 
edges represent pairwise relationships between objects (e.g., friendships, hyperlinks).
A consistent observation across different types of networks is that they are often composed of communities, i.e., groups of densely interconnected nodes \cite{Fortunato2010}.
A community is often associated with a group of nodes sharing a role or similarity such as 
a circle of friends in social networks \cite{Newman2004}, a set of web pages discussing the same topic \cite{Adamic2005,Karrer2011} and a functional group of proteins \cite{Jonsson2006}. 
 
Core-periphery structure is another mesoscopic structure of networks that has experienced a surge of interests in the last two decades.
A core-periphery structure in its simplest form refers to a partition of a network into two groups of nodes called core and periphery, where core nodes are densely interconnected (i.e., adjacent), and peripheral nodes are adjacent to the core nodes but not to other peripheral nodes \cite{Borgatti2000,Csermely2013,Rombach2017}.
Core-periphery structure has been detected in a number of networks including social networks \cite{Borgatti2000,Holme2005,Boyd2006,Rossa2013,Lee2014,Yang2014,Zhang2015,Cucuringu2016,Gamble2016,Kojaku2017,Ma2017,Rombach2017}, protein-protein interaction networks \cite{DaSilva2008,Yang2014,Bruckner2015}, neural networks \cite{Bassett2013,Tunc2015}, trade networks \cite{Boyd2010,Rossa2013,Ma2015}, financial networks \cite{Craig2014,IntVeld2014,Lee2014,Fricke2015,Barucca2016,Sardana2016} and transportation networks \cite{Rossa2013,Lee2014,Xiang2016,Kojaku2017,Rombach2017}.
For example, in a world-trade network among countries, economically strong countries trade with other strong countries, constituting a core. 
Economically weak countries mainly trade with strong countries, constituting a periphery \cite{Rossa2013,Ma2015}.

\begin{figure}
	\centering
	\includegraphics[width=0.5\hsize]{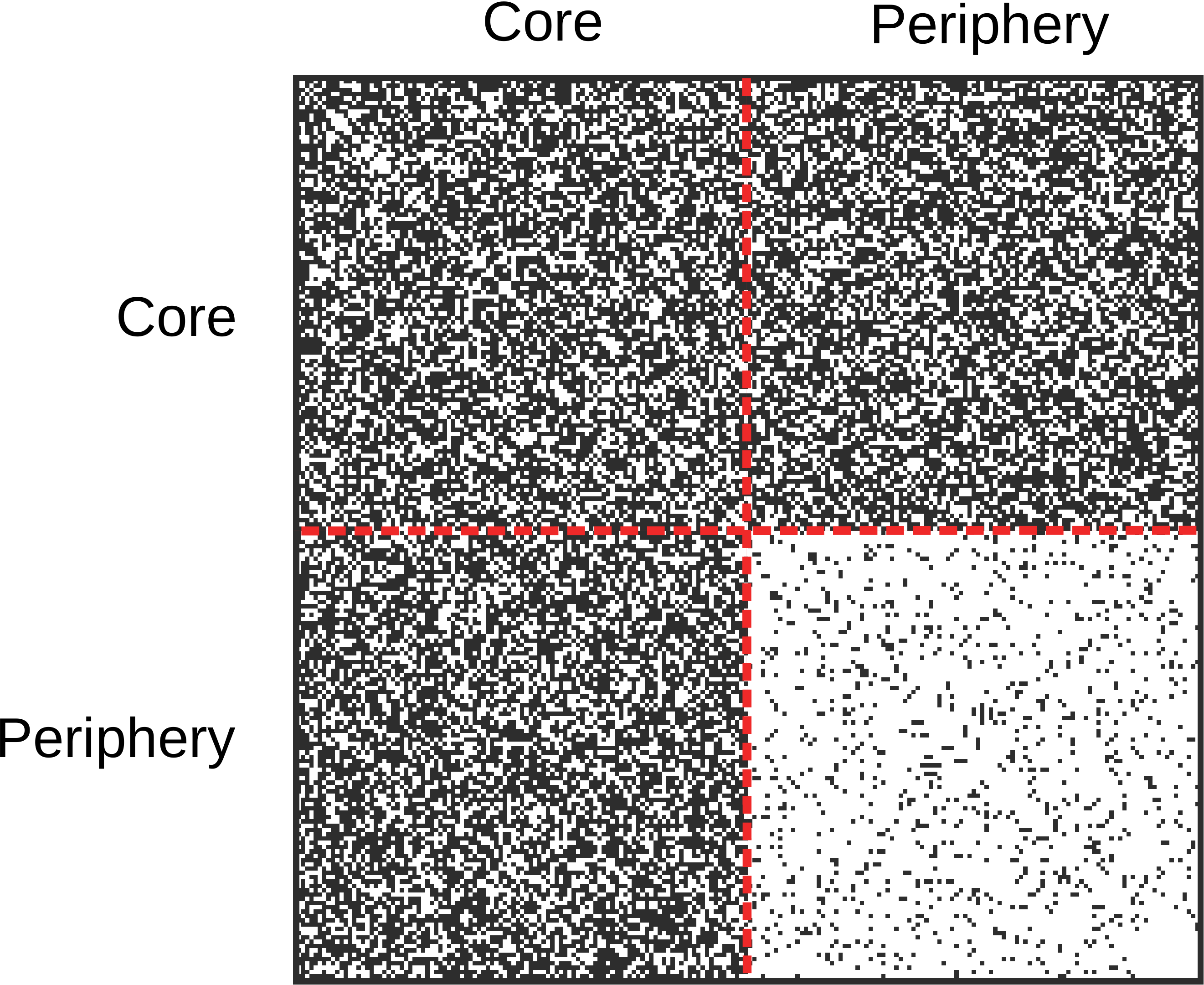}
	\caption{
		Adjacency matrix of a network with good-looking core-periphery structure composed of two blocks. 
		The filled and open cells in the $i$th row and $j$th column indicate the presence and absence of the edge between nodes $i$ and $j$, respectively.
		The dotted lines represent the boundary between the core and periphery.
	}
	\label{fig:single_cp}
\end{figure}

Borgatti and Everett analysed core-periphery structure in quantitative terms for the first time \cite{Borgatti2000}.
They expressed a core-periphery structure by a core block (i.e., group of core nodes) and a periphery block (i.e., group of peripheral nodes) as shown in Fig.~\ref{fig:single_cp}.
The core block has many intra-block edges (the top left block in Fig~\ref{fig:single_cp}). 
The periphery block has relatively few intra-block edges (the bottom right block in Fig.~\ref{fig:single_cp}).
There may be many inter-block edges (off-diagonal blocks in Fig.~\ref{fig:single_cp}) \cite{Borgatti2000,Boyd2006,Yang2014,Tunc2015,Kojaku2017,Ma2017,Rombach2017} or relatively few inter-block edges \cite{Borgatti2000,Boyd2010,Lip2011,Zhang2015,Ma2015,Cucuringu2016,Yan2016,Sardana2016,Fu2017,Rombach2017}.
The core-periphery structure expressed by blocks of nodes is classified as a discrete variant of core-periphery structure based on edge density \cite{Borgatti2000,Boyd2006,Lip2011,Csermely2013,Lee2014,Ma2015,Cucuringu2016,Gamble2016,Fu2017,Kojaku2017,Ma2017,Rombach2017}. 
There are other types of core-periphery and related structure, such as continuous versions of core-periphery structure \cite{Borgatti2000,Boyd2010,Lee2014,Cucuringu2016,Rombach2017}, transport-based core-periphery structure \cite{DaSilva2008,Rossa2013,Lee2014,Cucuringu2016,Marc2016}, $k$-core \cite{Alvarez-Hamelin2005} and rich-clubs \cite{Zhou2004,Colizza2006}.

Given that block structure of networks, or equivalently, hard partitioning of the nodes into groups, has spurred many studies such as community detection \cite{Luxburg2007,Fortunato2010} and the inference of stochastic block models (SBM) \cite{Karrer2011,Peixoto2017}, as well as its appeal to intuition, 
we focus on the discrete version of core-periphery structure based on edge density in the present paper. 
If a network has such core-periphery structure, the core block should have more intra-block edges and the periphery block should have fewer intra-block edges than a reference.
We argue that the core-periphery structure that Borgatti and Everett proposed (Fig.~\ref{fig:single_cp}), which many of the subsequent work is based on, is impossible if we use the configuration model \cite{Fosdick2016} as the null model and there are just one core and one periphery.
The configuration model is a common class of random graph models that preserve the degree or its mean value of each node. 
Therefore, our claim implies that there is no core-periphery structure a la mode de Borgatti and Everett beyond the expectation from the degree of each node (i.e., hubs are core nodes), which is, in fact, consistent with some previous observations \cite{IntVeld2014,Zhang2015,Barucca2016}.

Then, we are led to a question: what is a core-periphery structure?
To answer this question, let us look at the status of the configuration model in other measurements of networks.
We have a plethora of centrality measures for nodes because the degree is often not a useful measure of the importance of nodes \cite{Newman2010}.
In other words, different centrality measures provide rank orders of nodes in the given network that are not expected from the configuration model.
In network motif analysis, where one looks for small subnetworks that are abundant in a given network, we discount the frequency of subnetworks that are merely explained by the degree of the nodes (i.e., configuration model) \cite{Milo2002}. 
In community detection, it is conventional to use the configuration model as the null model against which one assesses the significance of community structure \cite{Newman2004,Fortunato2010,Karrer2011,Peixoto2017}. 
To solve the conundrum that one does not discover core-periphery structure using the configuration model as the null model, we propose that one must add at least one different block apart from a core block and the corresponding periphery block for a network to have core-periphery structure that is consistent with Fig.~\ref{fig:single_cp}. 
Such blocks may be a community, sparsely connected part, a different core-periphery pair \cite{Xiang2016,Kojaku2017,Ma2017,Rombach2017}, a core that shares the periphery with the focal core-periphery pair \cite{Yan2016} and so forth.
Then, we propose a scalable algorithm to partition a network into multiple core-periphery pairs including community detection as special cases, aiming to detect core-periphery structure that is not merely explained by the degree of each node. 
Crucially, we use the configuration model as the null model, which is different from our previous algorithm \cite{Kojaku2017}. 

\section{Core-periphery structure needs at least three blocks}
\label{sec:blockmodels}
Consider an unweighted network composed of $N$ nodes and $M$ edges.
The $N\times N$ adjacency matrix of the network is denoted by $\Amat=(A_{ij})$, where $A_{ij}=1$ if nodes $i$ and $j$ ($\neq i$) are adjacent and $A_{ij}=0$ otherwise.
We assume that the network is undirected (i.e., $A_{ij}=A_{ji}$ for all $i \neq j$) and has no self-loops (i.e., $A_{ii}=0$ for all $i$).
Let $d_i$ be the number of edges incident to node $i$ (i.e., degree). 
As the null model of networks, we use the configuration model, i.e., a random network model preserving the degree of each node.
For the configuration model, we allow multi-edges (i.e., multiple edges between nodes) and self-loops for computational ease. 
In fact, multi-edges and self-loops change our quality function for finding core-periphery structure in the order of $1/N$, which is negligible if $N$ is large. 
We denote by $\Exp[\cdot]$ the expectation with respect to the configuration model.

\begin{figure}
	\centering
	\includegraphics[width=1.01\hsize]{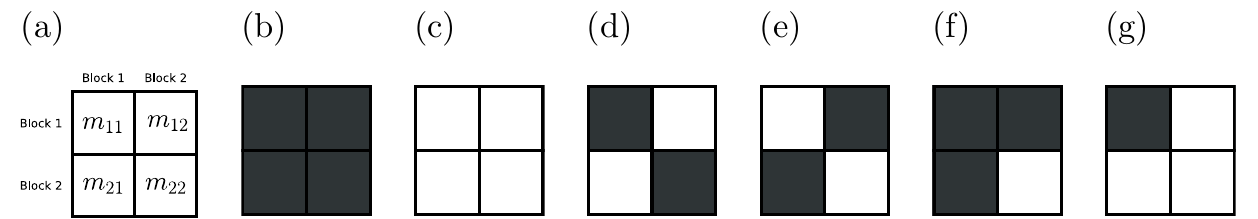}
	\caption{
		Schematic illustration of network structures composed of two blocks.
		The filled cells indicate that there are more edges than that expected for the configuration model (i.e., $m_{uv} > \Exp[m_{uv}]$).
		The open cells indicate that  there are fewer edges than that expected for the configuration model (i.e., $m_{uv} < \Exp[m_{uv}]$).
		Only the structures shown in (d) and (e) are possible.
	}
	\label{fig:blockmodel_bs2}
\end{figure}

Consider a partition of the set of $N$ nodes into $B$ blocks (i.e., groups).
Let $N_u$ be the number of nodes in block $u$ and $m_{uv}$ be the number of edges between blocks $u$ and $v$.
Note that $m_{uv} = m_{vu}$. 
For notational convenience, we define $m_{uu}$ as twice the number of self-loops in block $u$ plus twice the number of edges between different nodes within block $u$.
Denote by $m_{uv}^{\conf}$ the number of edges between blocks $u$ and $v$ in a network generated by the configuration model whose degree sequence is given by that of the original network.
Suppose a network composed of $B=2$ blocks (Fig.~\ref{fig:blockmodel_bs2}(a)).
There are potentially six types of block structure of networks represented by two blocks.
In Figs.~\ref{fig:blockmodel_bs2}(b)--\ref{fig:blockmodel_bs2}(g), 
a filled block has more edges than that for the configuration model (i.e., $m_{uv} > \Exp[m^{\conf} _{uv}]$), 
and a open block has fewer edges than that for the configuration model  (i.e., $m_{uv} < \Exp[m^{\conf} _{uv}]$). 
The entire network would be dense if there are many intra- and inter-block edges (Fig.~\ref{fig:blockmodel_bs2}(b)).
In contrast, the network would be sparse if there are relatively few intra- and inter-block edges (Fig.~\ref{fig:blockmodel_bs2}(c)). 
The network has community structure if there are many intra-block edges and relatively few inter-block edges (Fig.~\ref{fig:blockmodel_bs2}(d)).
A contrasting case is a structure close to a bipartite network, where there are relatively few intra-block edges and many inter-block edges (Fig.~\ref{fig:blockmodel_bs2}(e)).  
Core-periphery structure would correspond to the case in which there are many edges within one block and few edges within the other block.
With core-periphery structure, inter-block edges may be abundant (Fig.~\ref{fig:blockmodel_bs2}(f)) \cite{Borgatti2000,Boyd2006,Yang2014,Tunc2015,Kojaku2017,Ma2017,Rombach2017} or not (Fig.~\ref{fig:blockmodel_bs2}(g)) \cite{Borgatti2000,Boyd2010,Lip2011,Zhang2015,Ma2015,Cucuringu2016,Yan2016,Sardana2016,Fu2017,Rombach2017}.

\begin{figure}
\centering
\includegraphics[width=0.5\hsize]{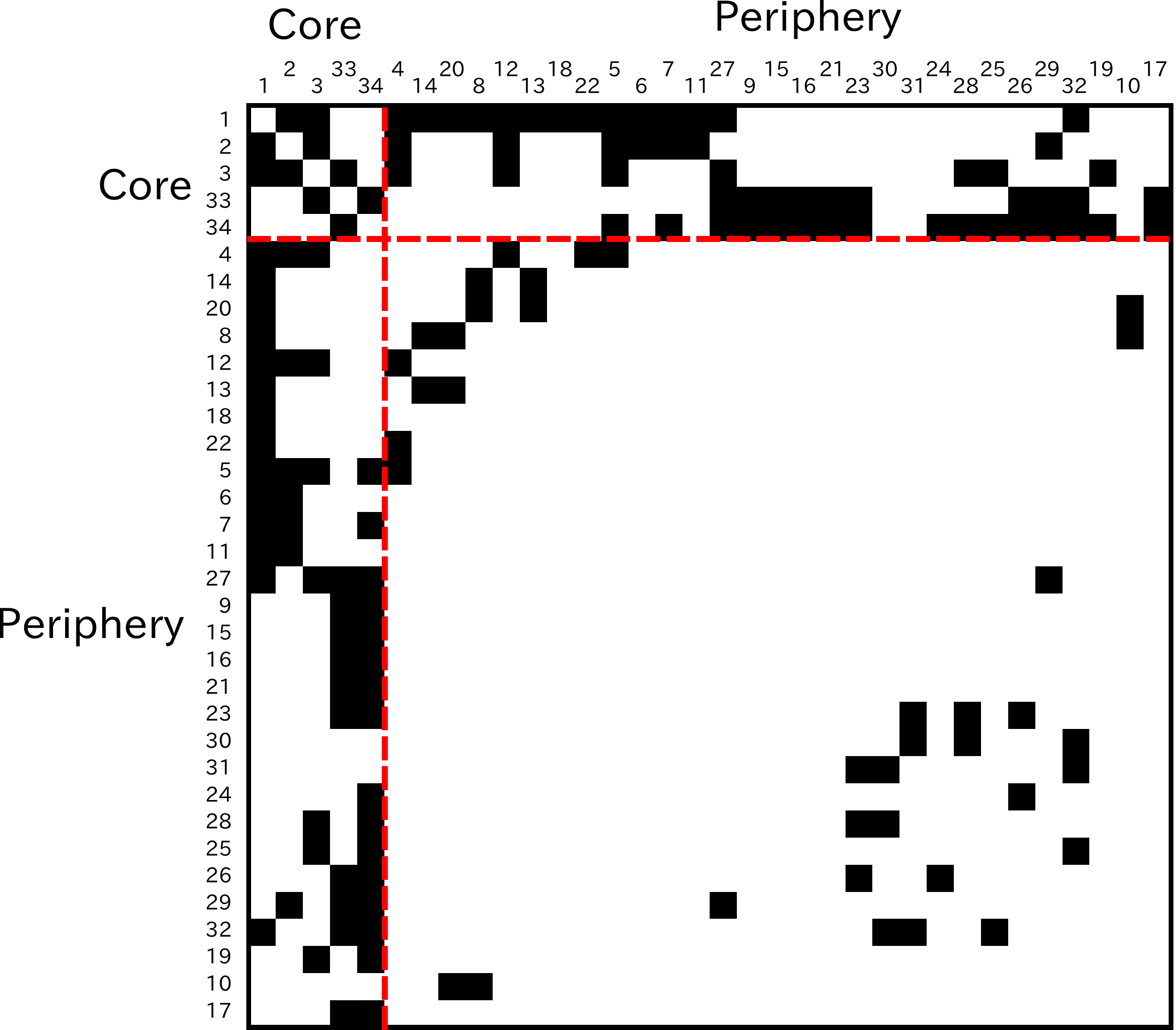}
\caption{
Core-periphery structure of the karate club network detected by the BE algorithm.
The nodes are reordered. 
The filled and open cells represent the presence and absence of edges, respectively. 
}
\label{fig:demo}
\end{figure}

Many algorithms for finding discrete versions of core-periphery structure seek a partition of nodes into one core block and one periphery block (Figs.~\ref{fig:blockmodel_bs2}(f) or \ref{fig:blockmodel_bs2}(g)).
Let us consider the karate club network \cite{Zachary1977}, which has been demonstrated to have core-periphery structure \cite{Rossa2013,Ma2015,Xiang2016,Fu2017,Kojaku2017,Ma2017,Rombach2017}.
The Borgatti-Everett (BE) algorithm partitions the $N=34$ nodes into a core and a periphery as shown in Fig.~\ref{fig:demo}. 
The detected blocks seem to suggest core-periphery structure because the core nodes are densely interconnected, whereas the peripheral nodes are sparsely interconnected.
However, relative to the configuration model, the network is closer to a bipartite network than to core-periphery structure; there are fewer edges within both core and periphery blocks (i.e., $m_{11} = 10$, $\Exp[m_{11} ^{\conf}] = 26.25$, $m_{22}=38$ and $\Exp[m^{\conf} _{22}]=54.25$) and more edges between the core and periphery blocks than those expected for the configuration model (i.e., $m_{12}= 54$ and $\Exp[m^{\conf} _{12}]=37.74$).%
 
This observation is in fact universal; core-periphery structure is impossible with two blocks when the null model is the configuration model.
To show this, consider a network composed of $B=2$ blocks.
The degree of each node is the same between the original network and a sample network generated by the configuration model. 
Therefore, the number of edges emanating from each block is also the same between the original network and the sample network. 
Therefore, we obtain
\begin{linenomath}
\begin{align}
	m_{11} + m_{12} &=  \Exp[ m^{\conf} _{11} ] + \Exp[ m^{\conf} _{12} ], \label{eq:m11} \\
	m_{21} + m_{22} &=  \Exp[ m^{\conf} _{21} ] + \Exp[ m^{\conf} _{22} ]. \label{eq:m22}
\end{align}
\end{linenomath}
Rearranging Eqs.~\eqref{eq:m11} and \eqref{eq:m22} yields
\begin{linenomath}
\begin{align}
	m_{11} -\Exp\left[ m^{\conf}_{11} \right] = -\left( m_{12} - \Exp\left[m^{\conf}_{12} \right] \right), \label{eq:block1}\\ 
	m_{22} -\Exp\left[ m^{\conf}_{22} \right] = -\left( m_{21} - \Exp\left[m^{\conf}_{21} \right] \right). \label{eq:block2}
\end{align}
\end{linenomath}
Equations~\eqref{eq:block1} and \eqref{eq:block2} imply that if a block has more intra-block edges in the original network than in the configuration model, 
the same block must have fewer inter-block edges in the original network than in the configuration model.
Because we assumed that the network is undirected, we obtain $m_{21} = m_{12}$ and $m^{\conf}_{21} = m^{\conf} _{12}$.
Using these relationships, we rewrite Eq.~\eqref{eq:block2} as 
\begin{linenomath}
\begin{align}
	m_{22} -\Exp\left[ m^{\conf}_{22} \right] = -\left( m_{12} - \Exp\left[m^{\conf}_{12} \right] \right). \label{eq:block22}
\end{align}
\end{linenomath}
By combining Eqs.~\eqref{eq:block1} and \eqref{eq:block22}, we obtain 
\begin{linenomath}
\begin{align}
	m_{11} - \Exp\left[ m^{\conf}_{11} \right] 
	= m_{22} - \Exp\left[ m^{\conf}_{22} \right]. \label{eq:intra-block}
\end{align}
\end{linenomath}
Equation~\eqref{eq:intra-block} indicates that there is no network composed of two blocks such that
the core block has more edges in the original network than the configuration model (i.e., $m_{11} > \Exp\left[ m^{\conf} _{11} \right]$) 
and the periphery block has fewer edges in the original network than the configuration model (i.e., $m_{22} < \Exp\left[ m^{\conf} _{22} \right]$).
Therefore, the core-periphery structure does not exist if one partitions a network into a single core block and a single periphery block, as the BE algorithm does. 
It should be noted that Eqs.~\eqref{eq:block1} and \eqref{eq:block2} imply that the networks represented by Figs.~\ref{fig:blockmodel_bs2}(b) and \ref{fig:blockmodel_bs2}(c) are also impossible. 
In contrast, the networks shown in Figs.~\ref{fig:blockmodel_bs2}(d) and \ref{fig:blockmodel_bs2}(e) satisfy Eqs.~\eqref{eq:block1}, \eqref{eq:block2}, and \eqref{eq:intra-block} and therefore are possible.

Core-periphery structure is possible if the network has $B=3$ or more blocks. 
To identify the block structures that are possible and those that are not, we introduce the notion of compatibility of block structure as follows.
Consider a network composed of $B$ blocks.
The number of edges between blocks $u$ and $v$ in the original network is given by
\begin{linenomath}
\begin{align}
	m_{uv} &= \sum_{i=1}^N \sum_{j=1}^N A_{ij} \delta(b_i,u) \delta(b_j,v), \label{eq:muv} 
\end{align}
\end{linenomath}
where $b_i$ is the index of the block to which node $i$ belongs, and $\delta(\cdot, \cdot)$ is Kronecker delta.
Equation~\eqref{eq:muv} leads to 
\begin{linenomath}
\begin{align}
	\sum_{v=1}^B m_{uv} &= \sum_{v=1} ^B \sum_{i=1}^N \sum_{j=1}^N A_{ij} \delta(b_i,u)\delta(b_j,v) \nonumber \\ 
	            &= \sum_{i=1}^N \delta(b_i,u) \sum_{j=1}^N A_{ij} \sum_{v=1}^B \delta(b_j,v) \nonumber \\ 
	            &= \sum_{i=1}^N \delta(b_i,u) \sum_{j=1}^N A_{ij} \nonumber \\ 
	            &= \sum_{i=1}^N  d_{i}\delta(b_i,u), \quad u=1,2, \ldots, B, \label{eq:m1}
\end{align}
\end{linenomath}
where $d_i=\sum_{j=1}^N A_{ij}$ is the degree of node $i$.
The sum $\sum_{v=1}^B m_{uv}$ is the sum of the degree of nodes in block $u$.
Because the configuration model preserves the degree of each node, the sum $\sum_{v=1}^B m_{uv}$ is the same between the original network and the configuration model, i.e., 
\begin{linenomath}
\begin{align}
	\sum_{v=1}^B m_{uv} = \sum_{v=1}^B \Exp\left[ m^{\text{conf}} _{uv} \right], \quad u=1, 2, \ldots, B. \label{eq:balance} 
\end{align}
\end{linenomath}
Note that Eq.~\eqref{eq:balance} generalises Eqs.~\eqref{eq:m11} and \eqref{eq:m22}. 
Then, we categorise blocks into dense (i.e., $m_{uv} > \Exp[m^{\conf} _{uv}]$) and sparse (i.e., $m_{uv} < \Exp[m^{\conf} _{uv}]$) blocks. 
We say that a block structure is compatible if the designated dense and sparse blocks are realisable in the sense that Eq.~\eqref{eq:balance} is satisfied.  
We describe the procedures to find compatible block structures in Appendix \ref{sec:compatibility}.

\begin{figure}
	\centering
	\includegraphics[width=1.01\hsize]{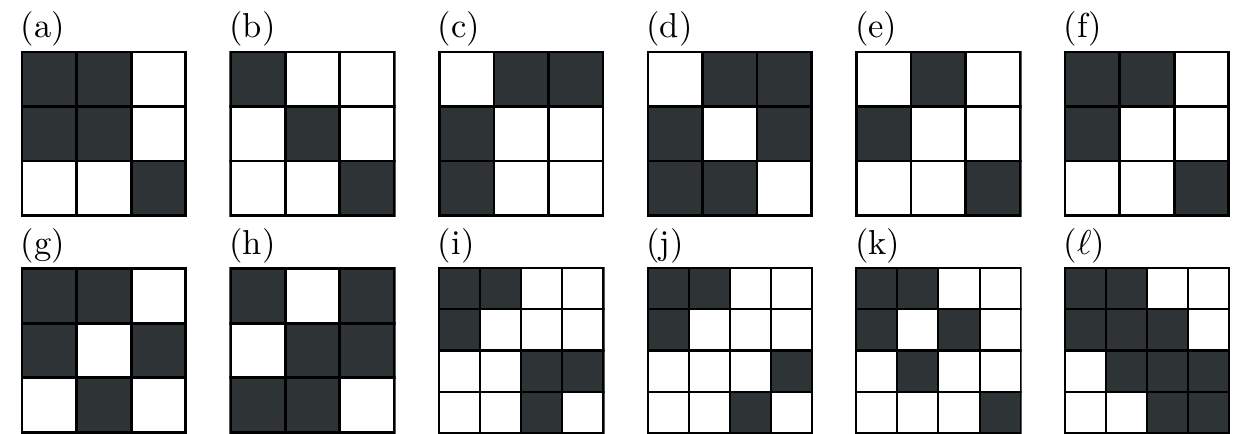}
	\caption{
		Schematic illustration of network structures with three or four blocks that are compatible with Eq.~(\ref{eq:balance}).
		We show all network structures composed of three blocks in panels (a)--(h) and four out of the 49 structures with four blocks in panels (i)--($\ell$).  
		The other 45 patterns with four blocks are shown in Fig.~\ref{fig:blockmodel_all}.
	}
	\label{fig:blockmodel_bs3}
\end{figure}

With $B=3$ blocks, eight types of block structure are compatible with Eq.~\eqref{eq:balance} (Figs.~\ref{fig:blockmodel_bs3}(a)--\ref{fig:blockmodel_bs3}(h)).
The networks shown in Figs.~\ref{fig:blockmodel_bs3}(a) and \ref{fig:blockmodel_bs3}(b) consist of two and three communities, respectively. 
The networks shown in Figs.~\ref{fig:blockmodel_bs3}(c) and \ref{fig:blockmodel_bs3}(d) are bipartite-like and tripartite-like networks, respectively.
The network shown in Fig.~\ref{fig:blockmodel_bs3}(e) is a union of a bipartite-like subnetwork composed of blocks 1 and 2 and a community composed of block 3.
These network structures extend those viable in the case of two blocks (Figs.~\ref{fig:blockmodel_bs2}(d) and \ref{fig:blockmodel_bs2}(e)).
The networks shown in Figs.~\ref{fig:blockmodel_bs3}(f)--\ref{fig:blockmodel_bs3}(h) contain core-periphery pairs.
In Fig.~\ref{fig:blockmodel_bs3}(f), blocks 1 and 2 constitute a core-periphery pair, and block 3 constitutes a community.
In Fig.~\ref{fig:blockmodel_bs3}(g), blocks 1 and 2 constitute a core-periphery pair, and blocks 2 and 3 constitute a bipartite-like subnetwork. 
The network shown in Fig.~\ref{fig:blockmodel_bs3}(h) consists of two cores (i.e., blocks 1 and 2) sharing a periphery (i.e., block 3), which is the structure studied in Ref.~\cite{Yan2016}.

\begin{figure}
	\centering
	\includegraphics[width=\hsize]{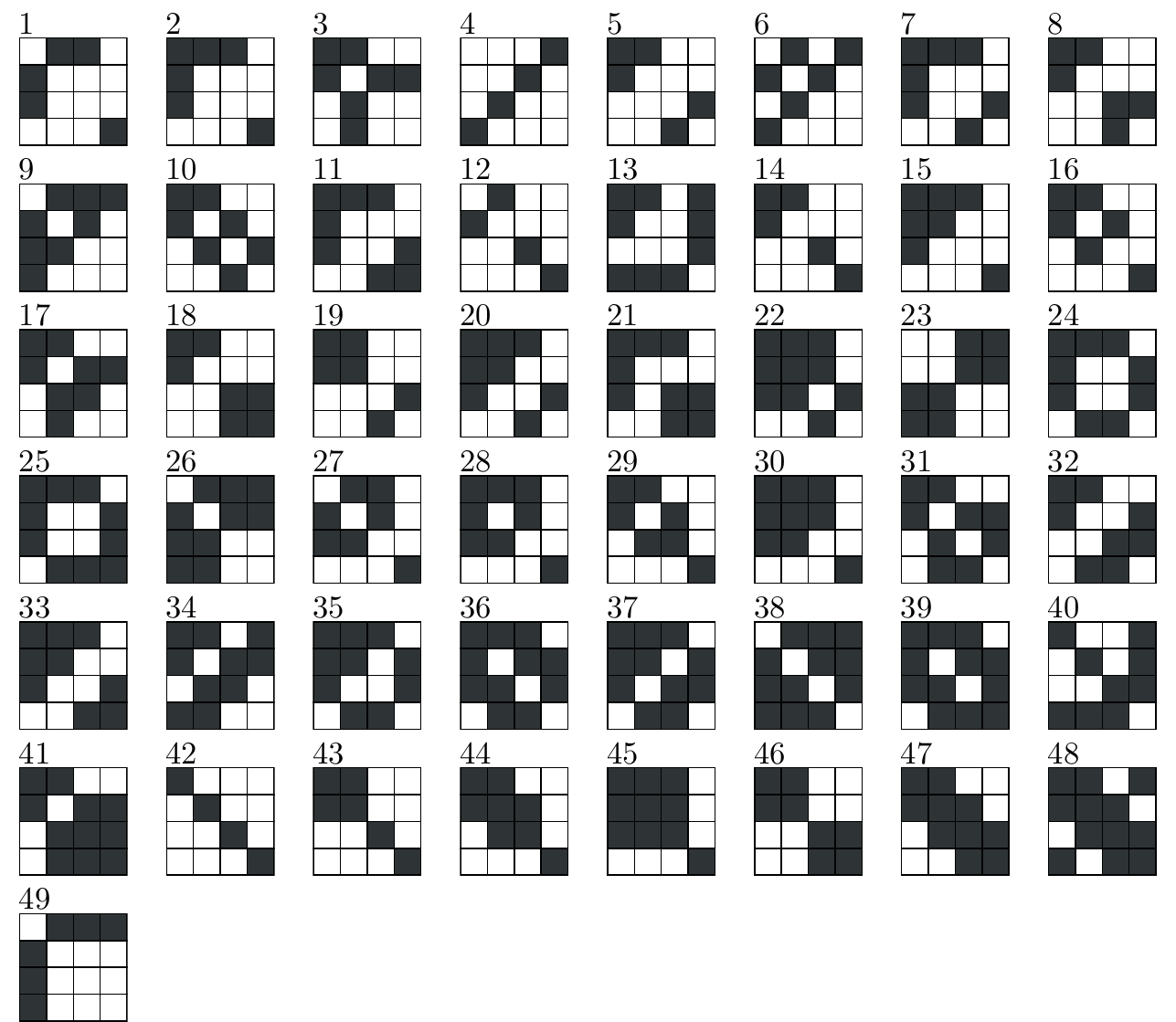}
	\caption{
		Schematic illustration of all network structures with four blocks that are compatible with Eq.~(\ref{eq:balance}).
		Patterns 5, 8, 16 and 47 are the same as those shown in Fig.~\ref{fig:blockmodel_bs3}. 
	}
	\label{fig:blockmodel_all}
	
\end{figure}

With $B=4$ blocks, 49 types of block structure are compatible with Eq.~\eqref{eq:balance}.
Four of them are shown in Figs.~\ref{fig:blockmodel_bs3}(i)--\ref{fig:blockmodel_bs3}($\ell$) for illustration (see Fig.~\ref{fig:blockmodel_all} for the others). 
The network shown in Fig.~\ref{fig:blockmodel_bs3}(i) is composed of two non-overlapping core-periphery pairs \cite{Tunc2015,Xiang2016,Kojaku2017,Ma2017}.
The network shown in Fig.~\ref{fig:blockmodel_bs3}(j) consists of one core-periphery pair (i.e., blocks 1 and 2) and one bipartite-like subnetwork (i.e., blocks 3 and 4). 
The network shown in Fig.~\ref{fig:blockmodel_bs3}(k) consists of one core-periphery pair (i.e., blocks 1 and 2), one bipartite-like subnetwork (i.e., blocks 2 and 3) and a community (i.e., block 4), in which the core-periphery pair and bipartite-like subnetwork overlap.
The network shown in Fig.~\ref{fig:blockmodel_bs3}($\ell$) has three overlapping communities, i.e., a community composed of blocks 1 and 2, one composed of blocks 2 and 3, and one composed of blocks 3 and 4. 

To conclude, the core-periphery structure a la mode de Borgatti and Everett \cite{Borgatti2000} relative to the configuration model can exist only when we have at least three blocks. 
In other words, a core-periphery pair requires a different substructure of the network that coexists in the same network, e.g., a community, bipartite-like structure, or another core-periphery pair that may overlap with the first one.

\section{Methods}
In this section, we first describe a new algorithm for detecting core-periphery structure, which we refer to as KM--config, based on the observations made in Section \ref{sec:blockmodels}. 
MATLAB and C$++$ codes of KM--config are available at \href{https://github.com/skojaku/km_config/}{\textcolor{blue}{https://github.com/skojaku/km\textunderscore{}config/}}.
Then, we explain other methods and data used in Section \ref{sec:results}.

\subsection{Our algorithm}
\subsubsection{Objective function}
We propose an algorithm, KM--config, to detect discrete versions of core-periphery structure in networks. 
In contrast to our previous algorithm that uses the \ER random graph as the null model \cite{Kojaku2017}, which we refer to as KM--ER, here we use the configuration model as the null model.
This is because we are interested in the structure that is not merely explained by the node's degree.

We assume that a network consists of $C$ non-overlapping core-periphery pairs, each of which 
is composed of one core block and one periphery block, e.g., Fig.~\ref{fig:blockmodel_bs3}(i).
Each core-periphery pair should have 
(i) many intra-core edges, 
(ii) many edges between the core and the corresponding periphery (i.e., core-periphery edges), 
(iii) few intra-periphery edges and 
(iv) few edges to other core-periphery pairs (i.e., inter-pair edges). 
Although some previous studies do not assume property (ii) \cite{Borgatti2000,Boyd2010,Lip2011,Craig2014,IntVeld2014,Fricke2015,Sardana2016,Xiang2016,Fu2017}, 
we require it because otherwise one cannot relate a periphery with a particular core.

We define idealised core-periphery pairs satisfying properties (i)--(iv) \cite{Tunc2015,Kojaku2017} by 
\begin{linenomath}
\begin{align}
	\label{eq:idealcp}
	\Amat^*=(A^* _{ij}),\quad A^* _{ij} \equiv (x_i + x_j - x_i x_j) \delta(c_i,c_j), 
\end{align}
\end{linenomath}
where $x_i=1$ or $x_i=0$ if node $i$ is a core node or a peripheral node, respectively, and $c_i$ ($1\leq c_i \leq C$) is the index of the core-periphery pair to which node $i$ belongs. 
Within each idealised core-periphery pair, every core node is adjacent to every other core node (property (i)) and also adjacent to all the corresponding peripheral nodes (property (ii)), and every peripheral node is not adjacent to any other peripheral nodes (property (iii)). 
Furthermore, there are no edges between different idealised core-periphery pairs (property (iv)).

We seek $c_i$ and $x_i$ ($1\leq i\leq N$) that maximise similarity between $\Amat$ and $\Amat^*$ as defined by
\begin{linenomath}
\begin{align}
	\label{eq:1}
	\Qcp \equiv& \frac{1}{2M}\sum_{i=1}^N \sum_{j=1}^N A_{ij}A^* _{ij} - \Exp\left[\frac{1}{2M}\sum_{i=1}^N \sum_{j=1}^N A^{\conf} _{ij}A^* _{ij}\right],
\end{align}
\end{linenomath}
where $\Amat ^{\conf} =(A^{\conf}_{ij})$ is the adjacency matrix of a network generated by the configuration model.
The first term on the right-hand side of Eq.~\eqref{eq:1} is the fraction of intra-core and core-periphery edges (i.e., $A_{ij}=A^{*} _{ij}=1$), corresponding to properties (i) and (ii).
The second term is the counterpart for the configuration model. 
The factor $1/2M$ in the first and second terms normalises $\Qcp$ to range in $[-1,1]$.
The remaining two properties (iii) and (iv) are also consistent with the maximisation of $\Qcp$. 
To show this, we rewrite $\Qcp$ as
\begin{linenomath}
\begin{align}
	\label{eq:qmin}
	\Qcp = - \frac{1}{2M}\sum_{i=1}^N \sum_{j=1}^N A_{ij}(1-A^* _{ij}) + \Exp\left[\frac{1}{2M} \sum_{i=1}^N \sum_{j=1}^N A_{ij} ^{\conf}(1-A^* _{ij})\right]. 
\end{align}
\end{linenomath}
Because $\sum_{i=1}^N \sum_{j=1}^N A_{ij}(1-A^* _{ij})$ is the sum of the number of intra-periphery edges and that of inter-pair edges, the maximisation of $\Qcp$ minimises the two types of edges associated with properties (iii) and (iv).

In the configuration model, the expected number of edges between nodes $i$ and $j$ is given by $\Exp[A_{ij} ^{\conf}]=d_id_j/2M$ \cite{Newman2003,Newman2006}.
Substitution of $\Exp[A_{ij} ^{\conf}]=d_id_j/2M$ and Eq.~\eqref{eq:idealcp} into Eq.~\eqref{eq:1} yields
\begin{linenomath}
\begin{align}
	\label{eq:q}
	\Qcp = \frac{1}{2M}\sum_{i=1}^N \sum_{j=1}^N \left( A_{ij} -\frac{d_id_j}{2M} \right) (x_i + x_j -x_i x_j )\delta(c_i,c_j).
\end{align}
\end{linenomath}
If we restrict that all nodes are core nodes (i.e., $x_i=1$ for $i=1,2,\ldots,N$), $\Qcp$ is equivalent to the modularity \cite{Newman2004,Newman2006}, which is used for finding communities in networks.
The $\Qcp$ shares shortcomings with the modularity such as the resolution limit. See Section \ref{sec:discussion} for further discussion.

\subsubsection{Relationship to Markov stability}
We can relate $\Qcp$ to discrete-time random walks, similar to the case of the Markov stability formalism for community detection \cite{Delvenne2010,Mucha2010,Lambiotte2014,Masuda2017}.
Consider a random walker that moves from a node to one of the neighbouring nodes selected uniformly at random in each discrete time step.
Let $T_{(c,x)(c',x')} \equiv m_{(c,x)(c',x')}/D_{(c,x)}$ be the transition probability from block $(c,x)$ to block $(c',x')$, where
$D_{(c,x)}$ is the sum of the degree of the nodes in block $(c,x)$. 
Let $\pi_{(c,x)} \equiv D_{(c,x)}/2M$ be the stationary probability with which the random walker visits block $(c,x)$.
Then, one can rewrite $\Qcp$ as 
\begin{linenomath}
\begin{align} 
    \label{eq:qrandomwalk}
		\Qcp &= \frac{1}{2M}\sum_{c=1}^C \left( 
		m_{(c,1)(c,1)} + 2m_{(c,0)(c,1)}
		-\frac{D_{(c,1)}^2}{2M}
		-\frac{2D_{(c,0)}D_{(c,1)}}{2M}
		\right) \nonumber \\
		&= \sum_{c=1}^C \left( 
		\frac{D_{(c,1)}}{2M} \cdot \frac{ m_{(c,1)(c,1)}}{D_{(c,1)}} 
		+\frac{2D_{(c,0)}}{2M} \cdot \frac{ m_{(c,0)(c,1)}}{D_{(c,0)}} 
		-\frac{D_{(c,1)}^2}{4M^2}
		-\frac{2D_{(c,0)}}{2M} \cdot \frac{D_{(c,1)}}{2M} 
		\right) \nonumber \\
		&= \sum_{c=1}^C \left( 
		\pi_{(c,1)} T_{(c,1)(c,1)} + 
		2\pi_{(c,0)} T_{(c,0)(c,1)} 
		-\pi_{(c,1)}^2 - 2\pi_{(c,0)}\pi_{(c,1)}
		\right). 
\end{align}
\end{linenomath}
Now, imagine a random walker starting from a node $i$ selected randomly according to the stationary density $d_i/2M$ ($1 \leq i\leq N$), at time $t=0$.
The probability that the random walker is in block $(c,x)$ at time $t=0$ and block $(c',x')$ at time $t=1$ is given by $\pi_{(c,x)}T_{(c,x)(c',x')}$, which is accounted for by the first and second terms of the right-hand side of Eq.~\eqref{eq:qrandomwalk}.
The corresponding probability for the configuration model is given by $\pi_{(c,x)}\pi_{(c',x')}$, which is accounted for by the third and fourth terms.
Therefore, $\Qcp$ measures how likely a random walker moves to the core of the currently visited node in one step relative to the probability expected for the configuration model.
This observation is exploited in a different algorithm to detect core-periphery structure of networks \cite{Rossa2013}.

\subsubsection{Maximisation of the objective function}
We maximise $\Qcp$ using a label switching heuristic \cite{Raghavan2007,Blondel2008}, which we have employed in our previous algorithm, KM--ER, that uses the \ER random graph as the null model \cite{Kojaku2017}. First, we initialise the labels by $c_i=i$ and $x_i=1$ ($1 \leq i \leq N$).
Then, we update the label of each node as follows.
Suppose that node $i$ has a neighbour in a core-periphery pair $c'$.
We tentatively assign node $i$ to the core (i.e., $(c_i,x_i)=(c',1)$) and compute the new value of $\Qcp$.
We also tentatively assign node $i$ to the periphery (i.e., $(c_i,x_i)=(c',0)$) and compute $\Qcp$.
We perform the tentative assignments for all the core-periphery pairs to which any neighbour of node $i$ belongs.
If any tentative assignments do not raise $\Qcp$, we do not update $(c_i,x_i)$.
Otherwise, we update ($c_i$, $x_i$) to the tentative label (i.e., $(c', 0)$ or $(c', 1)$) giving the largest increment in $\Qcp$.
We inspect each node in a random order.
If no node has changed its label during the inspection of all the $N$ nodes, we stop updating the labels.
Otherwise, we draw a new random order and inspect each node according to the new random order. 
We run this algorithm ten times starting from the same initial condition and adopt the node labelling that realises the largest value of $\Qcp$.

The increment in $\Qcp$ caused by updating node $i$'s label from $(c,x)$ to $(c', x')$ is given by
\begin{linenomath}
\begin{align}
    \label{eq:dq}
	\frac{1}{M} 
	\left[
		\tilde d_{i, (c',1)} +  x' \tilde d_{i, (c',0)} 
		- d_i\frac{D_{(c',1)} +  x' D_{(c',0)}}{2M}
		- \frac{d_i ^2}{4M} \left( x'- 2(x + x'  - x x') \delta(c,c') \right)
	\right]
	\nonumber \\
	-
	\frac{1}{M} 
	\left[
		\tilde d_{i, (c,1)} +  x \tilde d_{i, (c,0)} 
		 - d_i\frac{D_{(c,1)} +  x D_{(c,0)}}{2M}
		 + \frac{d_i ^2}{4M} x
	\right],
\end{align}
\end{linenomath}
where $\tilde d_{i,(c,x)}=\sum_{j=1}^N A_{ij}\delta(c_j,c)\delta(x_j,x)$ is the number of edges connecting node $i$ and block $(c, x)$. 
When inspecting node $i$, we calculate Eq.~\eqref{eq:dq} at most $2d_i$ times.
Therefore, the time needed for inspecting all nodes is ${\cal O}\left(\sum_{i=1}^N d_i\right)={\cal O}(M)$, and that of the entire algorithm is ${\cal O}(M\times \text{(the number of inspections over the $N$ nodes)})$.

\subsubsection{Statistical test}
\label{sec:statistical_test}
We define the quality $q$ of a core-periphery pair $c$ by its contribution to $\Qcp$, i.e.,  
\begin{linenomath}
\begin{align}
    \label{eq:qc}
	q \equiv \frac{1}{2M} \sum_{i=1}^N\sum_{j=1}^N \left(A_{ij}-\frac{d_id_j}{2M}\right)(x_i + x_j - x_ix_j)\delta(c_i,c)\delta(c_j,c).
\end{align}
\end{linenomath}
One may deem that a core-periphery pair is significant if its $q$ is statistically larger than the value expected for the configuration model.
However, $q$ may depend on the size (i.e., the number of nodes) $n$ of the core-periphery pair, as is the case for the modularity \cite{Leskovec2010}. 

Inspired by these considerations, we carry out a statistical test of the detected core-periphery pairs as follows.
We generate 500 randomised networks for the given network using the configuration model. 
Then, we detect core-periphery pairs in each randomised network.
We compute the quality $\hat q$ and size $\hat n$ of each core-periphery pair detected in the randomised network.
On the basis of the samples of $\hat q$ and $\hat n$, we infer the joint probability distribution $P(\hat q, \hat n)$ using the Gaussian kernel density estimator \cite{Wand1993,Scott2012}.
Finally, we regard the core-periphery pair detected in the original network with a quality value of $q$ to be significant if $q$ is statistically larger than 
that of the core-periphery pair of the same size $n$ detected in the randomised networks, i.e., if $P( \hat q \geq q \given n ) \leq \alpha$, where $P$ is the probability and $\alpha$ is a significance level. (See Appendix \ref{sec:kde} for the computation of $P(\hat q \geq q \given \hat n)$.) 
We refer to the nodes that do not belong to any significant core-periphery pair as residual nodes.
 
Because we carry out the test for each core-periphery pair in the original network, we have to correct the significance level to suppress false positives due to multiple comparisons.
To this end, we adopt the {\v{S}}id{\'{a}}k correction \cite{Sidak1967}, with which we test each core-periphery pair in the original network at a significance level of $\alpha = 1-(1-\alpha')^{1/C}$,
where $\alpha'$ is the targeted significance. We set $\alpha'=0.05$.

Empirical networks often have core-periphery pairs that are substantially larger than any of those detected in the 500 randomised networks (Section~\ref{sec:quality_vs_size}).
It is unlikely that one finds core-periphery pairs of the same size in randomised networks even if more samples of randomised networks are generated. 
The kernel density estimator enables us to infer $P(\hat q \geq q \given n)$ for large core-periphery pairs in the original network based on the quality and size of smaller core-periphery pairs detected in randomised networks. 

Quality $q$ may be significantly large for bipartite-like pairs of blocks (Fig.~\ref{fig:blockmodel_bs2}(e)). 
Therefore, if our algorithm detects bipartite-like pairs of blocks, we manually mark them and distinguish them from the core-periphery pairs.
Specifically, we regard a detected pair of blocks as bipartite-like if it has fewer intra-core edges than expected for the configuration model (i.e., if $m_{(c,1),(c,1)} < \Exp[m_{(c,1),(c,1)}]$). 
Otherwise we regard it as a core-periphery pair.
Our algorithm did not find other types of block pairs (i.e., those shown in Figs.~\ref{fig:blockmodel_bs2}(b), \ref{fig:blockmodel_bs2}(c) and \ref{fig:blockmodel_bs2}(g)) for the networks examined in the following sections. 

\subsection{Other algorithms for comparison}
\label{sec:algorithms}
We compare the present algorithm, KM--config, with three algorithms for finding a single core-periphery pair, i.e., the BE \cite{Borgatti2000}, MINRES \cite{Boyd2010,Lip2011} and SBM \cite{Zhang2015} algorithms, and three algorithms for finding multiple core-periphery pairs, i.e., Xiang \cite{Xiang2016}, Divisive \cite{Kojaku2017} and KM--ER algorithms \cite{Kojaku2017}.
We ran the Tun$\c{c}$--Verma \cite{Tunc2015} algorithm but do not show the results because the Tun$\c{c}$--Verma algorithm did not find significant core-periphery pairs or did not terminate within 48 hours on our computer (Intel 2.6GHz Sandy Bridge processors and 4GB of memory).
It should be noted that none of these algorithms uses the configuration model as the null model. 

The BE, Divisive and KM--ER algorithms intend to produce many core-periphery edges (i.e., edges connecting a core node and a peripheral node) within each core-periphery pair (Fig.~\ref{fig:blockmodel_bs2}(f)).
With the MINRES, SBM and Xiang algorithms, core-periphery edges can be relatively sparse (Fig.~\ref{fig:blockmodel_bs2}(g)).

We set the parameters of these algorithms as follows.
For the SBM algorithm, we set $\gamma_k$, $p_{kl}$ ($1\leq k,l\leq 2$) in Ref.~\cite{Zhang2015} to $\gamma_1 = \gamma_2 = 0.5$, $p_{11}=0.5$, $p_{12}=p_{21} = \rho^2$ and $\rho_{22}=\rho^4$, where $\rho=2M/[N(N-1)]$.
The Xiang algorithm has a parameter, denoted by $\beta \in[0,1]$ in Ref.~\cite{Xiang2016}, to tune the number of core-periphery pairs.
We set to $\beta=1$.
The Xiang algorithm uses a centrality measure to find core-periphery pairs. 
Therefore, we adopt the degree centrality measure. 
Note that the authors of Ref.~\cite{Xiang2016} claim that the choice of the centrality measure does not considerably affect the results.
With the Xiang algorithm, each node may belong to multiple core-periphery pairs.
Therefore, if a node belongs to multiple core-periphery pairs, we assign the node to the core-periphery pair to which the extent of belonging is the largest.
If a node belongs to multiple core-periphery pairs to the same extent, then we assign the node to one of the core-periphery pairs selected with equal probability.
The other algorithms do not have parameters.
As is the case of KM--config, the BE, SBM, Divisive and KM--ER algorithms are stochastic.
Therefore, we run the BE, SBM, Divisive or KM--ER algorithm ten times and use the best core-periphery pairs in terms of the algorithm-specific quality function.

For the core-periphery pairs detected by the six previous algorithms, we carry out our previously proposed statistical test \cite{Kojaku2017} that adopts the \ER random graph model as the null model.
The statistical test runs as follows. 
Suppose that a network is composed of a single core-periphery pair.
We generate $500$ randomised networks using the \ER random graph with the same number of nodes and edges as the original network.
Then, we detect a single core-periphery pair in each of the randomised networks using the BE algorithm and compute its quality by 
\begin{linenomath}
\begin{align}
    \label{eq:bgquality}
    \frac{
        \sum_{i=1}^N\sum_{j=1}^{i-1}(A_{ij}-\rho)(A^* _{ij}-\rho^*)
    }{
        \sqrt{\sum_{i=1}^N\sum_{j=1}^{i-1}(A _{ij}-\rho)^2}\sqrt{ \sum_{i=1}^N\sum_{j=1}^{i-1}(A^* _{ij}-\rho^*)^2}
    },
\end{align}
\end{linenomath}
where $A^*$ is given by Eq.~\eqref{eq:idealcp} and $\rho^*=\sum_{i=1}^N \sum_{j=1}^{i-1}A^{*}_{ij}/[N(N-1)/2]$.
If the quality of the core-periphery pair detected in the original network is larger than a fraction $1-\alpha$ of those detected in the randomised networks, then we regard the core-periphery pair in the original network as significant. 
It should be noted that this test is not applicable when the null model is the configuration model.
If we use the configuration model as the null model,
any core-periphery pair detected in the original network will be judged to be insignificant because no network is partitioned into a single core-periphery pair whose $q$ value is larger than that for the configuration model. 

If we detect multiple core-periphery pairs in the original networks, we apply the same statistical test for each of them \cite{Kojaku2017}.
Specifically, for each core-periphery pair, we construct a subnetwork composed of the nodes and edges within the focal core-periphery pair.
Then, we apply the statistical test to the subnetwork.
We correct the significance level using the {\v{S}}id{\'{a}}k correction \cite{Sidak1967}; we test each core-periphery pair in the original network at a significance level of $\alpha = 1-(1-\alpha')^{1/C}$, where $\alpha'=0.05$ and $C$ is the number of core-periphery pairs detected in the original network.

\subsection{Data}
\label{sec:dataset}
We analyse the 12 empirical networks listed in Table~\ref{ta:net_stat}.
We discard the direction and weight of the edge.

In the karate club network, each node represents the member of a university's karate club \cite{Zachary1977}.
Two members are defined to be adjacent if they frequently interact outside the club activities. 
The club experienced a fissure as a result of a conflict between the instructor and the president.
Based on their self-reports, each node has a label indicating either the instructor's side (15 members),  president's side (16 members) or neutral (3 members).

In the dolphin social network, each node represents a dolphin living near Doubtful Sound in New Zealand \cite{Lusseau2003}.
An edge between two dolphins indicates that they were frequently observed in the same school during 1994 and 2001.
Each dolphin has a label indicating the sex, i.e., female (25 dolphins), male (33 dolphins) and unknown (4 dolphins). 

In the network of novel Les Mis\'{e}rables, each node is a character of the book \cite{Knuth1993}. 
Two characters are defined to be adjacent if they appear in the same chapter. 
The book consists of 365 chapters, most of which are a few pages long.

In the Enron email network, each node is an email account of the staff of Enron Inc \cite{Klimt2004}.
An edge indicates that an email was sent from one account to another account during the observation period.

In the jazz network, each node represents a jazz musician \cite{Gleiser2003}. 
Two jazz musicians are defined to be adjacent if they have played in the same band.

In the co-authorship network, 
each node represents a researcher in network science \cite{Newman2006}.
An edge indicates that two researchers have a joint paper. 
The nodes and edges were retrieved from all the references cited by two influential review papers on network science.
Then, the author of Ref.~\cite{Newman2006} manually added some nodes and edges and excluded those not belonging to the largest connected component. 

In the blog network, 
each node represents a blog on the United States presidential election in 2004 \cite{Adamic2005}.
Each edge indicates that one blog has a hyperlink to the other blog on its top page.
The blogs and their labels were collected from several blog directories \cite{Adamic2005}.
If a blog was unlabeled or had conflicting labels, the authors of Ref.~\cite{Adamic2005} manually determined the label. 
There are 586 liberal blogs and 636 conservative blogs.

In the worldwide airport network, each node is an airport \cite{Openflight.org,ToreOpsahl}.
An edge represents a direct commercial flight between two airports.
We use the network provided in Ref.~\cite{ToreOpsahl}.

In the protein-protein interaction network, each node is a human protein \cite{Rual2005}.
An edge indicates the presence of physical interaction between two proteins. 

In the network of chess players, each node represents a chess player \cite{KONECT}.
Two players are adjacent if they have played before.

In the co-authorship network of the arXiv astro-ph section, each node is a researcher \cite{Leskovec2007}. 
An edge indicates that two researchers have a joint paper in the arXiv's astro-ph section. 

In the network of the Internet, a node is an autonomous system (AS), i.e., a set of routers (or IP routing prefixes) managed by a network operator \cite{KONECT}.
An edge indicates a logical peering relationship between two ASes.

\section{Results}
\label{sec:results}
\subsection{Quality and size of detected core-periphery pairs}
\label{sec:quality_vs_size}
\begin{figure}
	\centering
	\includegraphics[width=1.01\hsize]{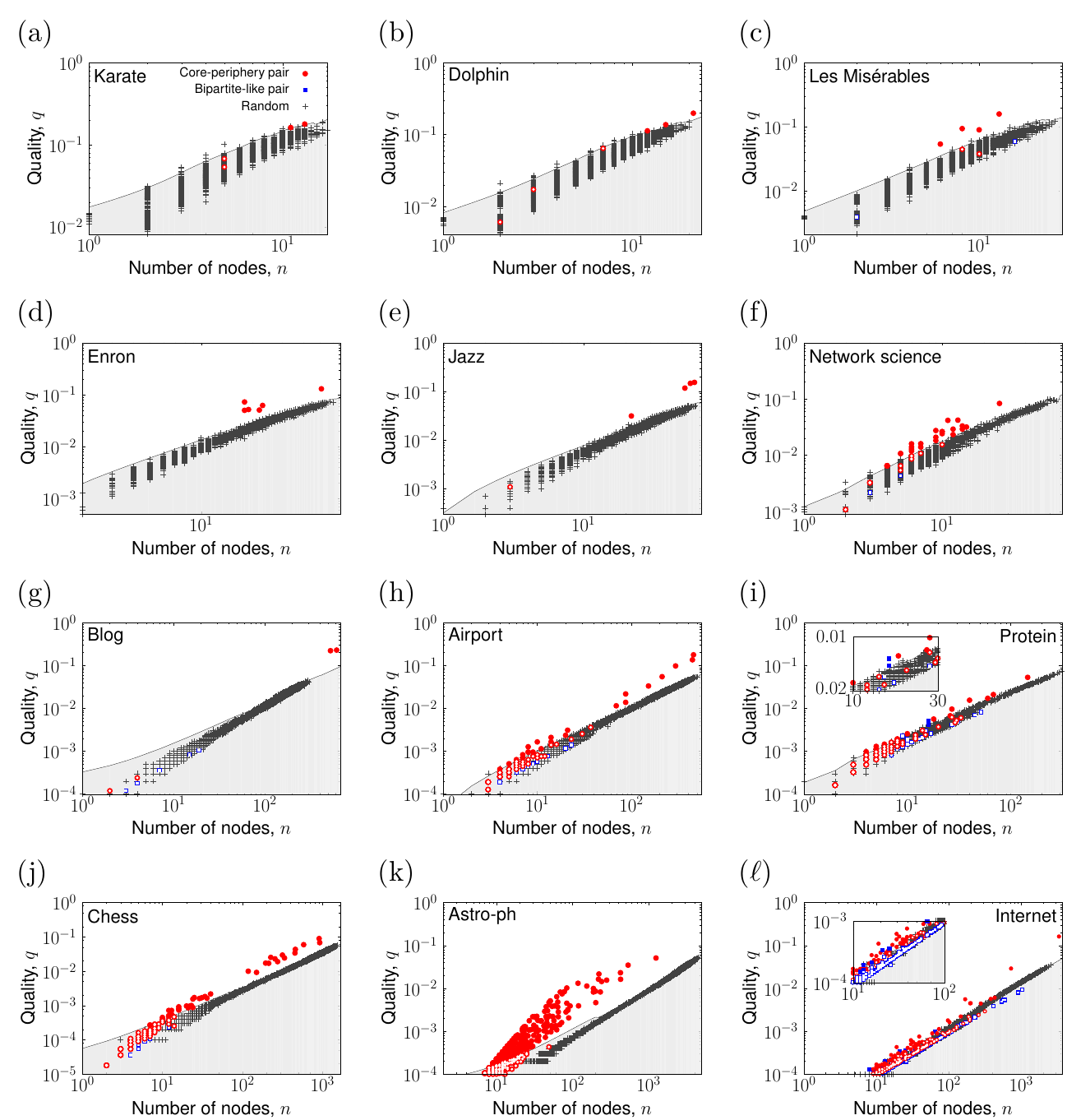}
	\caption{
		The quality of a core-periphery pair or bipartite-like pair, $q$, plotted against its number of nodes, $n$, in the 12 empirical networks and the corresponding randomised networks. 
		We used the KM--config algorithm. 
		The filled squares and filled circles represent significant core-periphery pairs and significant bipartite-like pairs, respectively.  
		In the shaded regions, the detected core-periphery and bipartite-like pairs are insignificant, as represented by open circles and open squares, respectively.  
		The crosses represent core-periphery or bipartite-like pairs detected in randomised networks.
		The insets in panels (i) and ($\ell$) magnify the region where many significant bipartite-like pairs lie.
	}
	\label{fig:qdist}
\end{figure}

The circles in Fig.~\ref{fig:qdist} represent the quality and size (defined as the number of nodes) of core-periphery pairs detected by KM--config in the 12 empirical networks. 
A larger core-periphery pair tends to have a large quality, $q$. This is also the case for the randomised networks (crosses in Fig.~\ref{fig:qdist}). 
Some core-periphery pairs detected in the empirical networks have a significantly larger $q$ value than those of the same size detected in the randomised networks. 
Our statistical test suggests that these core-periphery pairs are significant (circles outside the shaded regions in Fig.~\ref{fig:qdist}).
We find bipartite-like pairs in the 7 out of the 12 networks (squares in Fig.~\ref{fig:qdist}), some of which are significant in 2 out of the 7 networks (Figs.~\ref{fig:qdist}(i) and \ref{fig:qdist}($\ell$)). 
In 2 out of the 12 networks, we find significant core-periphery pairs that are larger than any of those detected in the corresponding randomised networks (Figs.~\ref{fig:qdist}(g) and \ref{fig:qdist}($\ell$)).  

\subsection{Core nodes are not necessarily hub nodes}
\label{sec:degree_vs_cp}
\begin{figure}
	\centering
	\includegraphics[width=1.01\hsize]{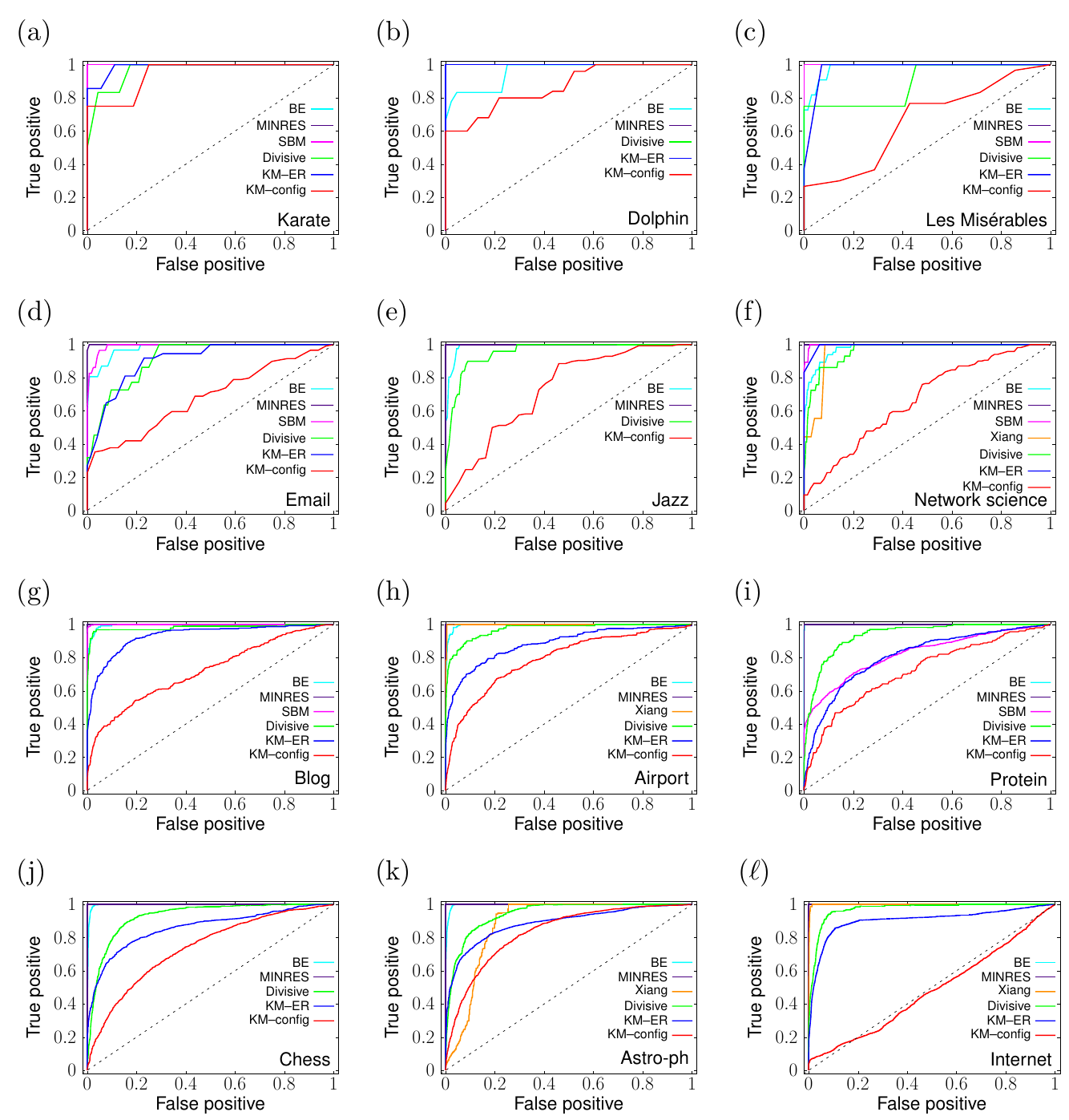}
	\caption{
	With KM--config, whether the node belongs to a core or periphery is not strongly associated with the node's degree.
	Each curve represents the relationship between the fraction of hub nodes in the set of significant core nodes (i.e., true positive rate) and that in the set of significant peripheral nodes (i.e., false positive rate).
        The dashed lines are the diagonal.
	The ROC curve is not shown if an algorithm does not detect any significant core-periphery pairs. 
	Some ROC curves perfectly overlap on top of each other, and this occurs if and only if the ROC curves pass through $(0, 1)$.
    	Most previous algorithms classify nodes into a core and periphery largely based on the degree of nodes.
	}
	\label{fig:roc}
\end{figure}

With KM--config, whether the node belongs to a core or periphery is not strongly associated with the node's degree.
To show this, we carry out a receiver operating characteristic (ROC) analysis (Fig.~\ref{fig:roc}). 
Let us regard $\theta N$ nodes ($\theta \in \{0,1/N,2/N,\ldots,1\}$) with the largest degree as hub nodes and the remaining nodes as non-hub nodes.
The ROC curves show the relationship between the fraction of hub nodes in the set of significant core nodes (i.e., true positive rate) and that in the set of significant peripheral nodes (i.e., false positive rate) when one varies the threshold $\theta$.
If all core nodes have a larger degree than all peripheral nodes, the ROC curve passes through $(0, 1)$ of the unit square (Fig.~\ref{fig:roc}).
If the degree of core nodes and that of peripheral nodes obey similar distributions, then the ROC curve is close to the diagonal line for the entire range of $\theta$. 

The area under the curve (AUC) of each ROC curve is shown in Table \ref{ta:auc}.
If the two distributions are completely separated, the AUC is equal to one. If they completely overlap, the AUC is equal to 0.5.
The AUC values for the BE, MINRES, SBM, Xiang and Divisive algorithms are fairly large (mostly above 0.95) for all the networks.
Therefore, these algorithms have a strong tendency to classify the nodes with a large degree as core nodes and those with a small degree as peripheral nodes.
The AUC values for KM--ER are also large (above 0.81) but not as large as those for the five algorithms.
Finally, KM--config determines the role (i.e., core or periphery) of each node by the degree of the node to the least extent, as suggested by the smallest AUC values across different networks among all the algorithms. 
These results on the AUC values are consistent with visual observations one can make in Fig.~\ref{fig:roc}.

\begin{figure}
	\centering
	\includegraphics[width=1.01\hsize]{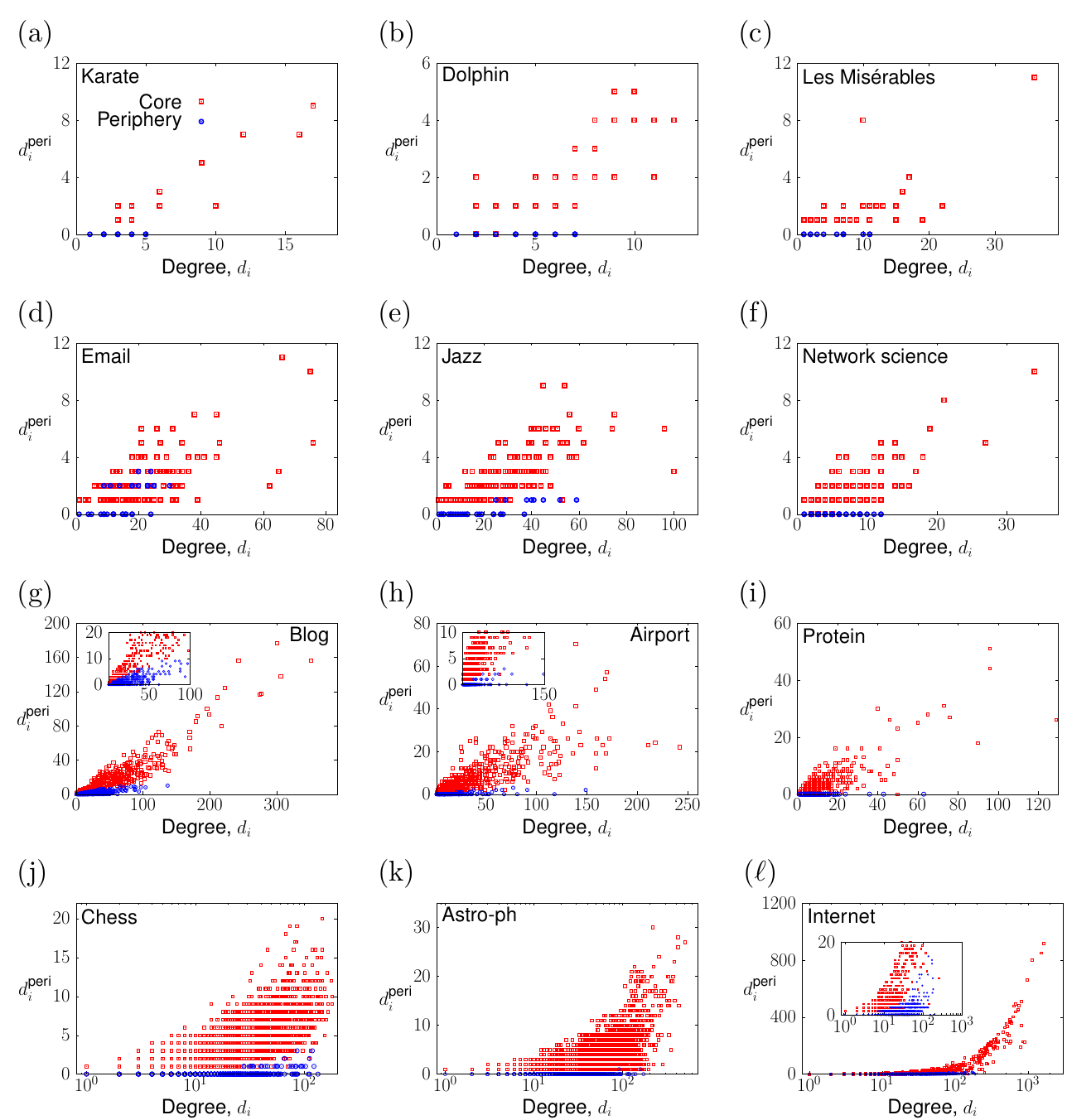}
	\caption{
		Relationships between $d_i$ and $d^{\text{peri}} _i$ for the empirical networks.
		The squares and circles indicate core nodes and peripheral nodes identified by KM--config, respectively.
		The insets of the panels (g), (h) and ($\ell$) magnify the regions with small $d_i ^{\text{peri}}$ values. 
	}
	\label{fig:neighbours}
\end{figure}

\begin{figure}
	\centering
	\includegraphics[width=1.01\hsize]{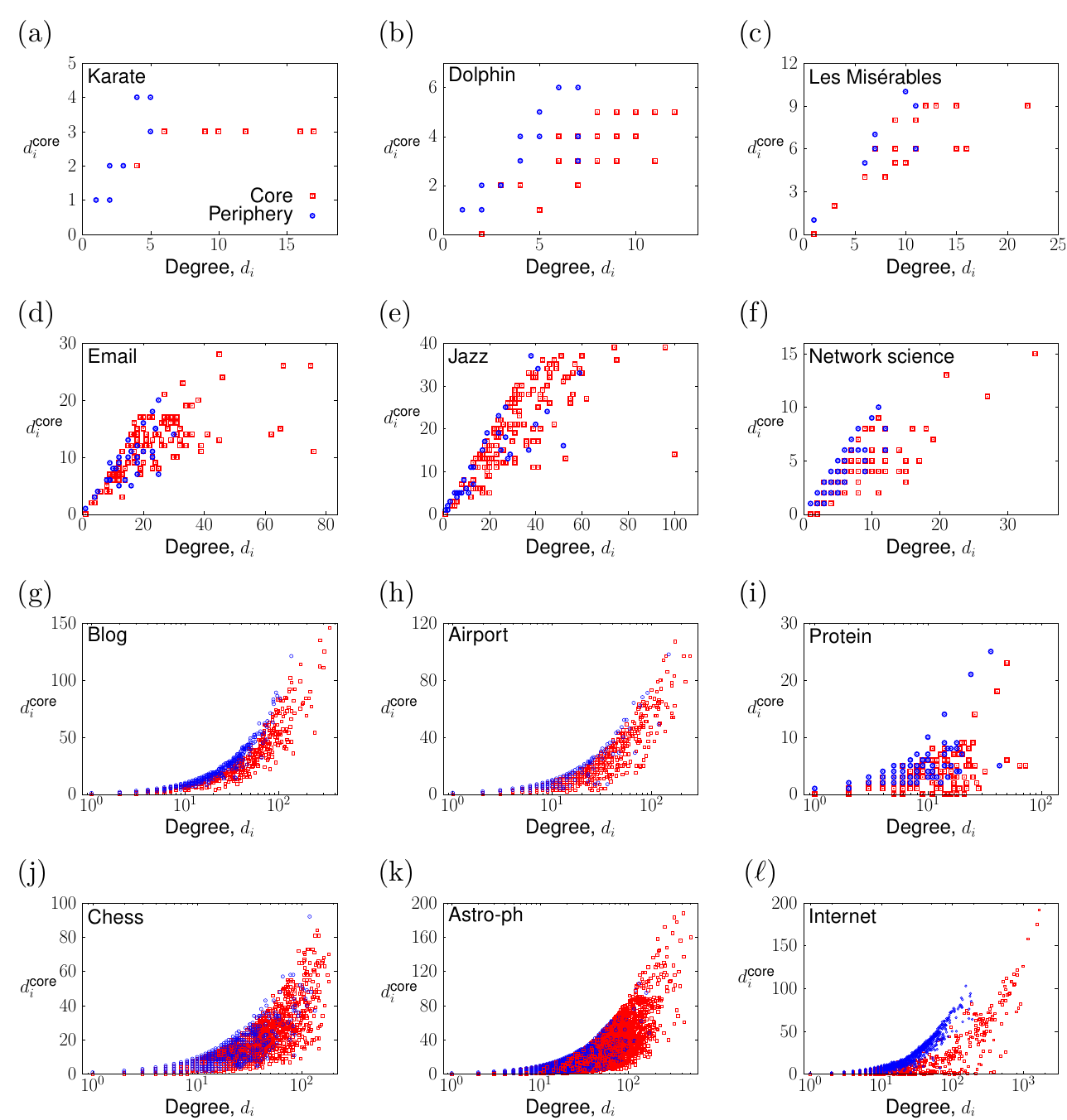}
	\caption{
		Relationships between $d_i$ and  $d^{\text{core}} _i$ for the empirical networks.
		The squares and circles indicate core nodes and peripheral nodes identified by KM--config, respectively.
	}
	\label{fig:neighbours_dcore}
\end{figure}

To illuminate on the meaning of the core and peripheral nodes detected by KM--config, 
let us denote by $d_i^{\text{core}}$ and $d_i^{\text{peri}}$ the number of neighbouring core and peripheral nodes of node $i$, respectively, within the core-periphery pair to which node $i$ belongs.
For each node $i$, we plot $d_i ^{\text{peri}}$ against degree $d_i$ in Fig.~\ref{fig:neighbours}. 
We find that peripheral nodes are adjacent to a smaller number of peripheral nodes within the same core-periphery pair (i.e., a small $d^{\text{peri}}_i$) than core nodes with a similar $d_i$ value would do.
This result is consistent with the concept of core-periphery structure based on edge density. 
However, this property is not necessarily respected if one classifies nodes according to the degree of each node.
In fact, with the other six algorithms, core nodes and peripheral nodes are less distinct from each other in terms of $d_i^{\text{peri}}$ (Figs.~\ref{fig:neighbours_be}--\ref{fig:neighbours_kmer} in Appendix \ref{sec:results_of_other_algorithms}).
As a corollary, with KM--config, the peripheral nodes tend to be more frequently connected to the core nodes within the same core-periphery pair than the core nodes do (Fig. \ref{fig:neighbours_dcore}). 
For some core nodes with a large degree, $d_i^{\text{core}}$ is equal to zero, which happens when a core-periphery pair has only one core node and forms a star.
 
\subsection{A core-periphery pair is a community?}
\label{sec:vscommunity}
We compare the core-periphery pairs identified by KM--config and communities in networks.
Here we determine communities by modularity maximisation using the Louvain algorithm \cite{Blondel2008}.
We run it ten times and adopt the node partition that realises the largest modularity value.
Table~\ref{ta:vscommunity} reports the modularity values for the node partition identified by the Louvain algorithm and that determined by KM--config, with the insignificant core-periphery pairs being included. 
The modularity value for the node partitioning into core-periphery pairs is close to 
that obtained by the modularity maximisation for most of the empirical networks.
Therefore, the detected core-periphery pairs may be similar to communities.

This result poses a question whether a core-periphery pair is a community in the traditional sense, and if so
whether the KM--config algorithm effectively classifies the nodes in each community into a core and a periphery according to the composition of intra- and inter-community edges that each node owns. 
To examine this point, we analyse the role of each node using a cartographic representation of networks \cite{Guimera2005a,Guimera2005}.
With the cartographic representation, the role of each node $i$ in a network is characterised by the standardised within-module degree $z_i \in [-\infty,\infty]$ and the participation coefficient $p_i \in [0,1]$ \cite{Guimera2005a,Guimera2005}.
They are defined by
\begin{linenomath}
\begin{align}
z_i &\equiv \frac{\displaystyle \tilde d_{i,c_i} - \langle \tilde d_{c_i} \rangle }{\displaystyle \sigma_{c_i}}, \\
p_i &\equiv 1 - \sum_{c=1} ^C \left( \frac{\tilde d_{i,c}}{d_i} \right)^2,
\end{align}
\end{linenomath}
where $\tilde d_{i,c}$ is the number of neighbours of node $i$ in the $c$th core-periphery pair ($1\leq c\leq C$), 
$c_i$ is the core-periphery pair to which node $i$ belongs,
$\langle \tilde d_{c_i} \rangle$ is the average of $\tilde d_{j,c_i}$ over the nodes $j$ in the  $c_i$th core-periphery pair including the case $j=i$, 
and $\sigma_{c_i}$ is the unbiased estimation of the standard deviation of $\tilde d_{j,c_i}$ over the nodes $j$ in the  $c_i$th core-periphery pair. 
A large $z_i$ value indicates that node $i$ has relatively many neighbours within the same core-periphery pair.
The $p_i$ value is the smallest if node $i$ is adjacent only to the nodes in a single core-periphery pair 
and largest if node $i$ is adjacent to an equal number of nodes across all core-periphery pairs.
In the cartographic representation of networks, each node $i$ is classified according to the position $(z_i,p_i)$ of the node in the $z$--$p$ space. 
The nodes are categorised into seven roles \cite{Guimera2005a,Guimera2005}. 
Here we do not use this categorisation rule but examine the distributions of the core and peripheral nodes in the $z$--$p$ space. 

\begin{figure}
	\centering
	\includegraphics[width=1.01\hsize]{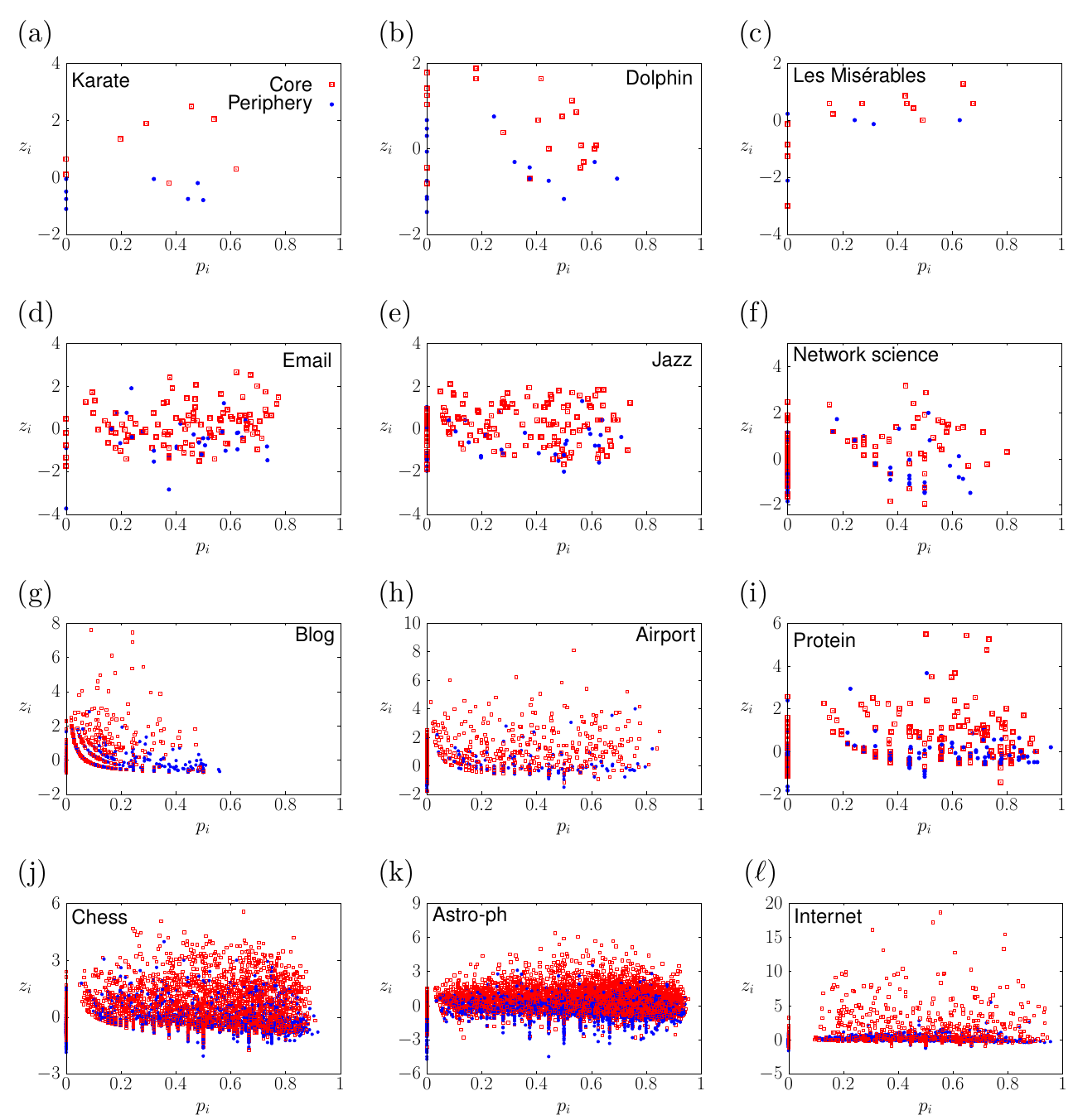}
	\caption{
		Cartographic analysis of the empirical networks.
		The squares and circles indicate core nodes and peripheral nodes identified by KM--config, respectively.
	}
	\label{fig:cartography}
\end{figure}

Figure~\ref{fig:cartography} shows $z_i$ and $p_i$ of each node for the 12 empirical networks. 
KM--config classifies the nodes having very large $z_i$ as core nodes.
However, for the other nodes, the values of $z_i$ and $p_i$ are not predictive of whether a node is in the core or periphery.
Therefore, the core and periphery that we propose are distinct from the roles of nodes identified by the cartographic analysis.

We find different results for the Divisive algorithm, which first divides the network into communities and then estimates the role of each node (i.e., core or periphery). 
With Divisive, the core nodes have larger $z_i$ than most of the peripheral nodes (Fig.~\ref{fig:cartography_dv} in Appendix~\ref{sec:results_of_other_algorithms}), indicating that the core nodes detected by Divisive largely correspond to the hub nodes as identified by the cartographic analysis.
This is because Divisive uses the BE algorithm to partition each community into a core and a periphery.
As shown in Fig.~\ref{fig:roc}, the BE algorithm classifies nodes into a core and a periphery by the degree of each node to a large extent.
Therefore, Divisive regards the nodes with a large $z_i$ as core nodes.
 
\subsection{Case studies}
In this section, we present case studies of some of the empirical networks analysed in the previous sections.

\begin{figure}
	\centering
	\includegraphics[width=1.01\hsize]{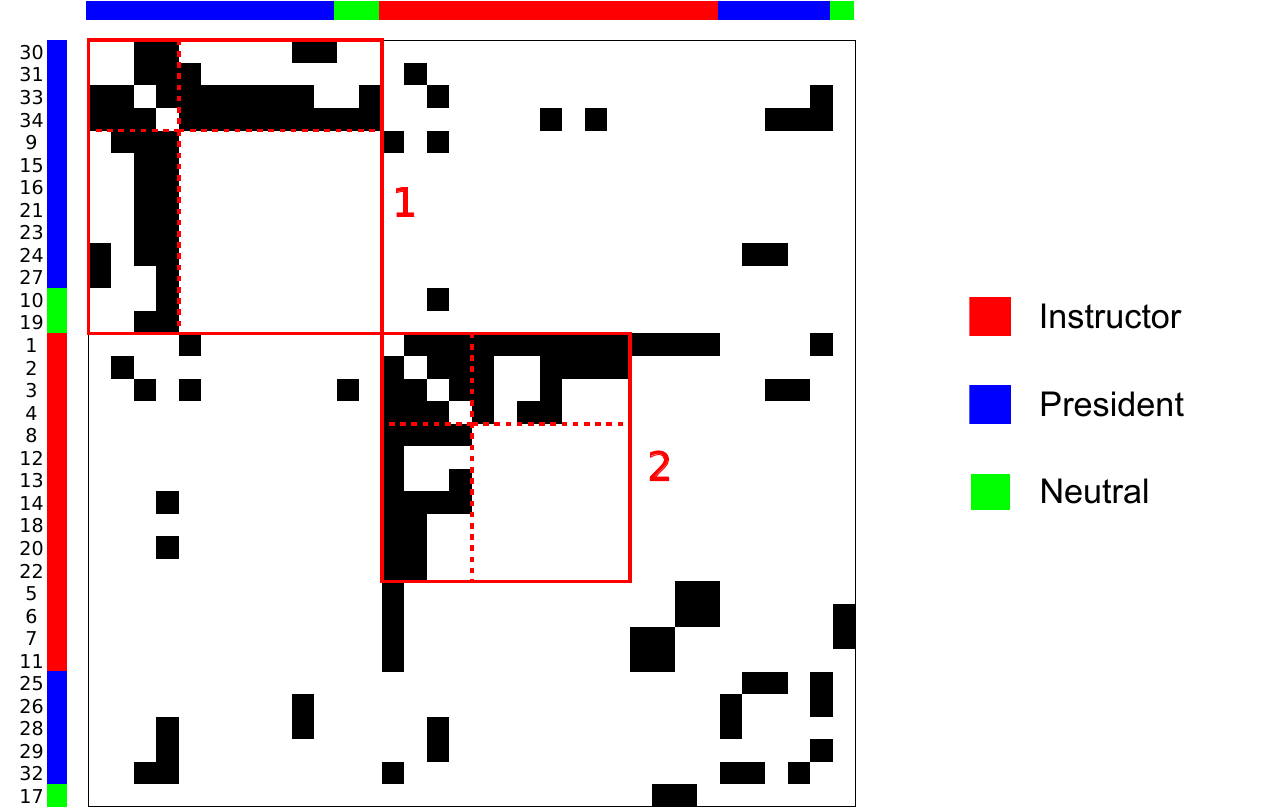}
	\caption{
		Significant core-periphery pairs in the karate club network.
		The rows and columns correspond to nodes.  
		The filled and open cells represent the presence and absence of an edge, respectively. 
		The square box bordered by the solid lines inside the adjacency matrix represents a significant core-periphery pair. 
		The dotted lines represent the boundary between the core and periphery in the core-periphery pair.
    		The colour of the node indicates the label of the member. 
	}
	\label{fig:karate}
\end{figure}

The core-periphery pairs in the karate club network are shown in Fig.~\ref{fig:karate}.
KM--config detected two significant core-periphery pairs and ten residual nodes.
A majority of the members on the president side  (12 members; 75\%), including the president (node 34), belong to core-periphery pair 1.
A majority of the members on the instructor side (11 members; 73\%), including the instructor (node 1), belong to core-periphery pair 2. 
These results are consistent with the social conflict of the club.
The residual consists of four members on the instructor side, five members on the president side and one neutral member. 
The significant core-periphery pairs are similar to those detected by our previous algorithm, KM--ER \cite{Kojaku2017}.

\begin{figure}
	\centering
	\includegraphics[width=1.01\hsize]{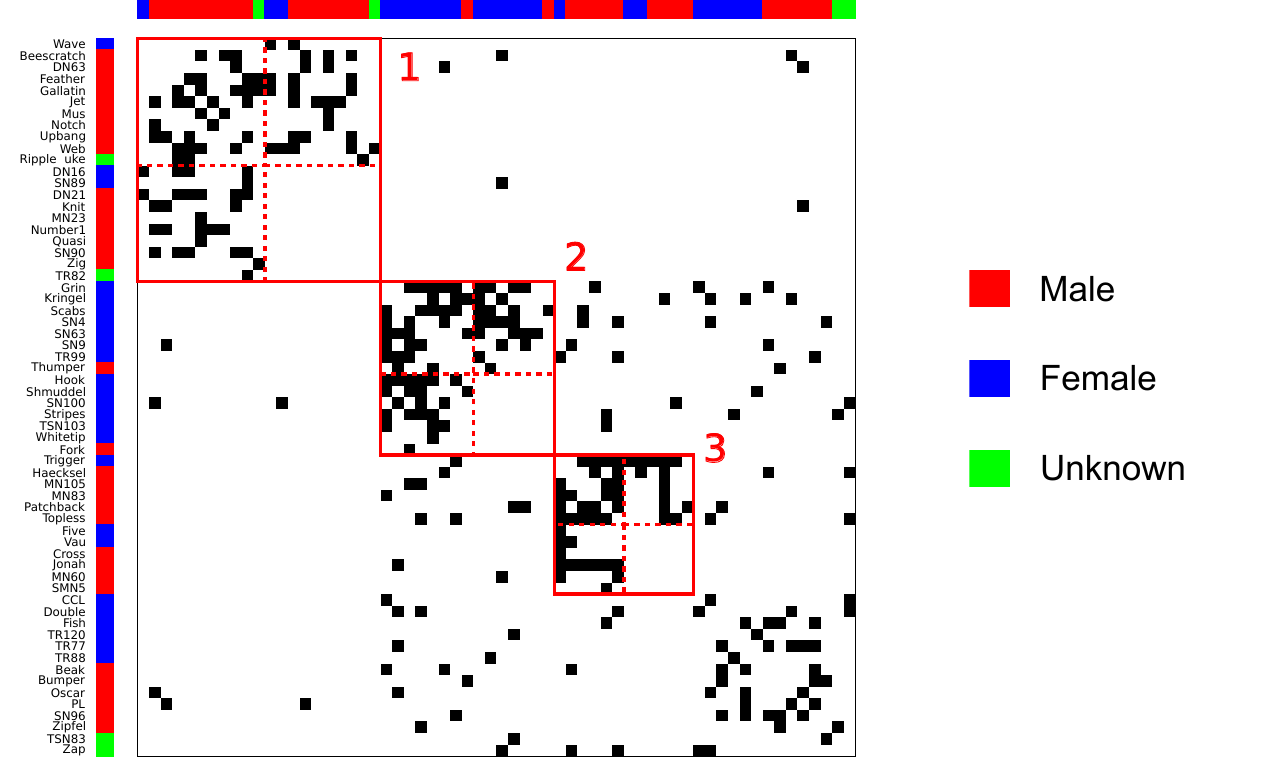}
	\caption{
		Significant core-periphery pairs in the dolphin social network.
    		The colour indicates the sex of the individual. 
	}
	\label{fig:dolphin}
\end{figure}

The core-periphery pairs in the dolphin social network are shown in Fig.~\ref{fig:dolphin}.
KM--config detected three significant core-periphery pairs and 14 residual nodes. 
Each core-periphery pair mostly consists of the dolphins of the same sex; there are two male-dominant core-periphery pairs (pairs 1 and 3) and one female-dominant core-periphery pair (pair 2).
A previous study identified five communities in the dolphin network by modularity maximisation \cite{Newman2004}, three of which are similar to the present core-periphery pairs 1, 2 and 3.

\begin{figure}
	\centering
	\includegraphics[width=0.6\hsize]{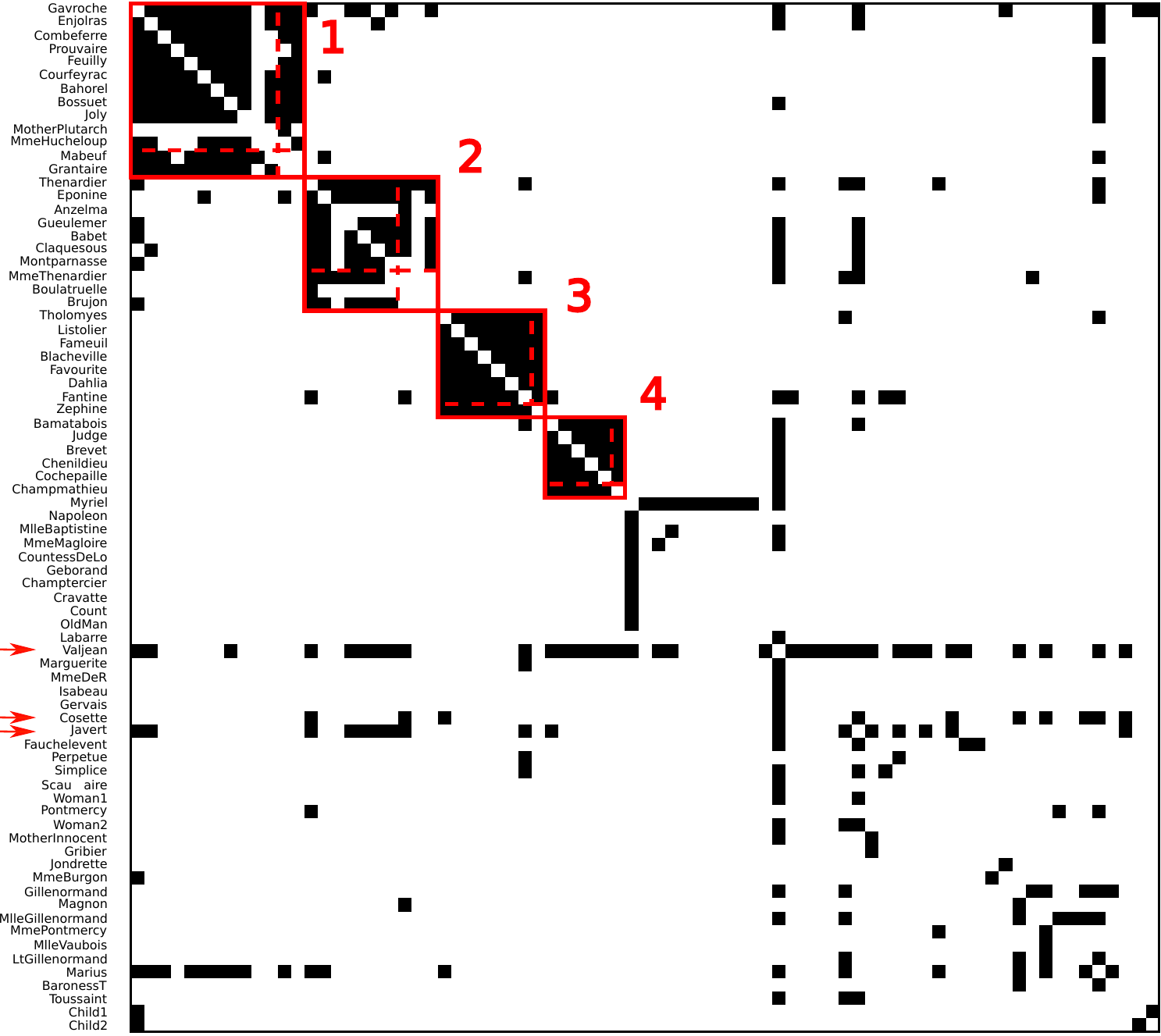}
	\caption{
		Significant core-periphery pairs in the network of Les Mis\'{e}rables.
		The arrows indicate main characters of the book.
	}
	\label{fig:lesmis}
\end{figure}

For the network of Les Mis\'{e}rables, KM--config identified four significant core-periphery pairs and 40 residual nodes (Fig.~\ref{fig:lesmis}).
A majority of nodes belonging to the significant core-periphery pairs are core nodes, suggesting that each core-periphery pair resembles a community.
In fact, a previous study used modularity maximisation to identify 11 communities in the same network \cite{Newman2004}, four of which are similar to our core-periphery pairs 1--4.
The significant core-periphery pairs are consistent with the plot of the story; 
the characters in core-periphery pairs 1, 2, 3 and 4 are the members of a revolutionary student club, 
 Th\'{e}nardier family and a street gang, Fantine's relatives and her friends, and characters involved in the Champmathieu's trial, respectively.
The main characters, e.g., Valjean, Javert and Cosette, are classified as residual nodes (arrows in Fig.~\ref{fig:lesmis}).
Although they have a large degree, they are regarded as residual nodes because they belong to insignificant core-periphery pairs. 

\begin{figure}
	\centering
	\includegraphics[width=1.01\hsize]{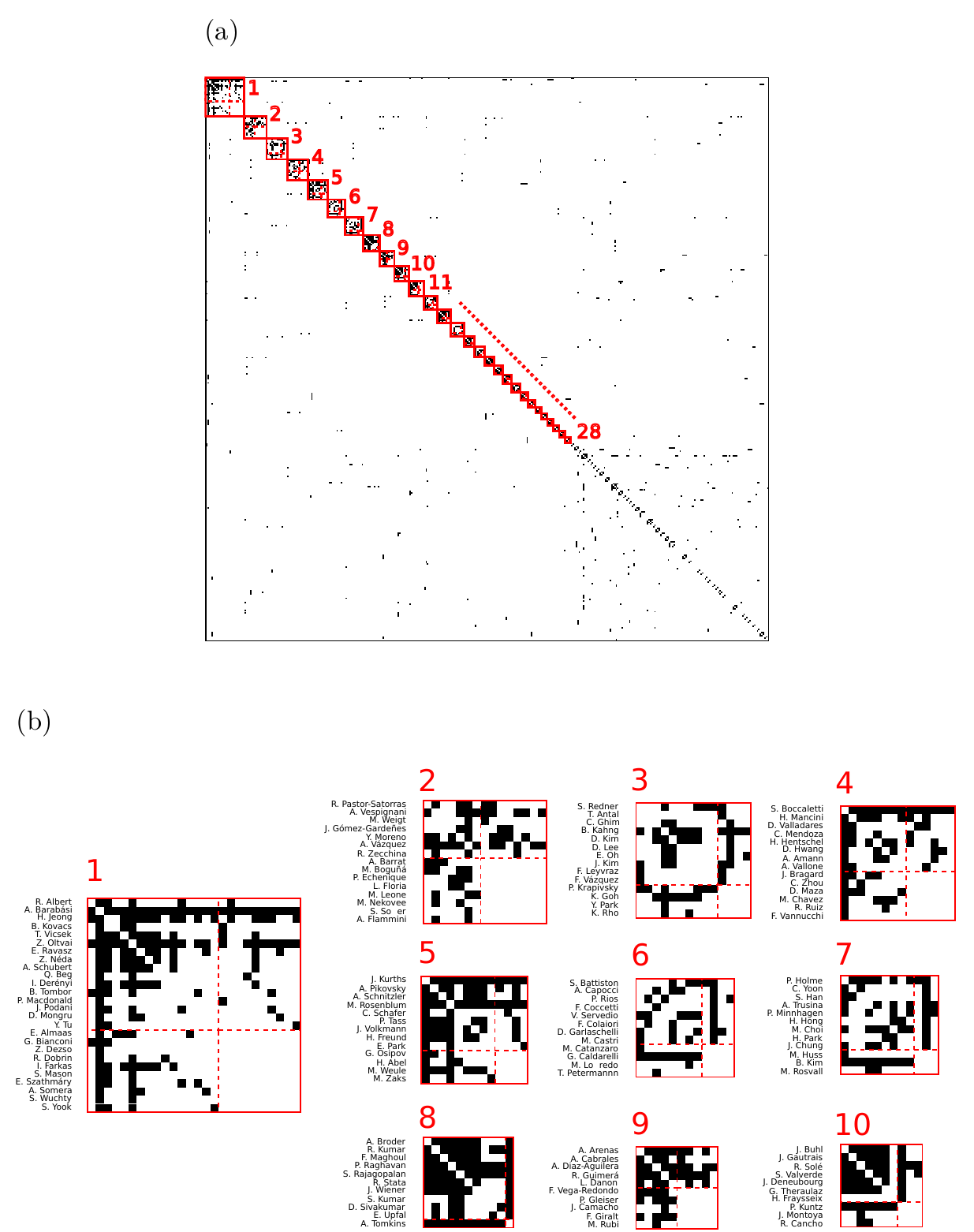}
	\caption{
		Significant core-periphery pairs in the co-authorship network in network science.
		(a) Adjacency matrix.
		(b) Structure of core-periphery pairs 1--10 in detail.
	}
	\label{fig:netscience}
\end{figure}

For the co-authorship network, KM--config detected 28 significant core-periphery pairs and 133 residual nodes (Fig.~\ref{fig:netscience}(a)).
Detailed structure of core-periphery pairs 1--10 is shown in Fig.~\ref{fig:netscience}(b).
Five core-periphery pairs (pairs 1, 5, 8, 9 and 10) have relatively many intra-core edges, many core-periphery edges and no intra-periphery edges, indicating a strong core-periphery structure.
Core-periphery pair 8 contains only one peripheral node, implying a structure close to a community.
Some core researchers collaborate with most of the researchers in the same core-periphery pair, e.g., 
A. Barab{\'a}si, H. Jeong and Z. Oltvai in core-periphery pair 1, A. V\'azquez and A. Vespignani in core-periphery pair 2 and S. Boccaletti in core-periphery pair 4.

\begin{figure}
	\centering
	\includegraphics[width=1.01\hsize]{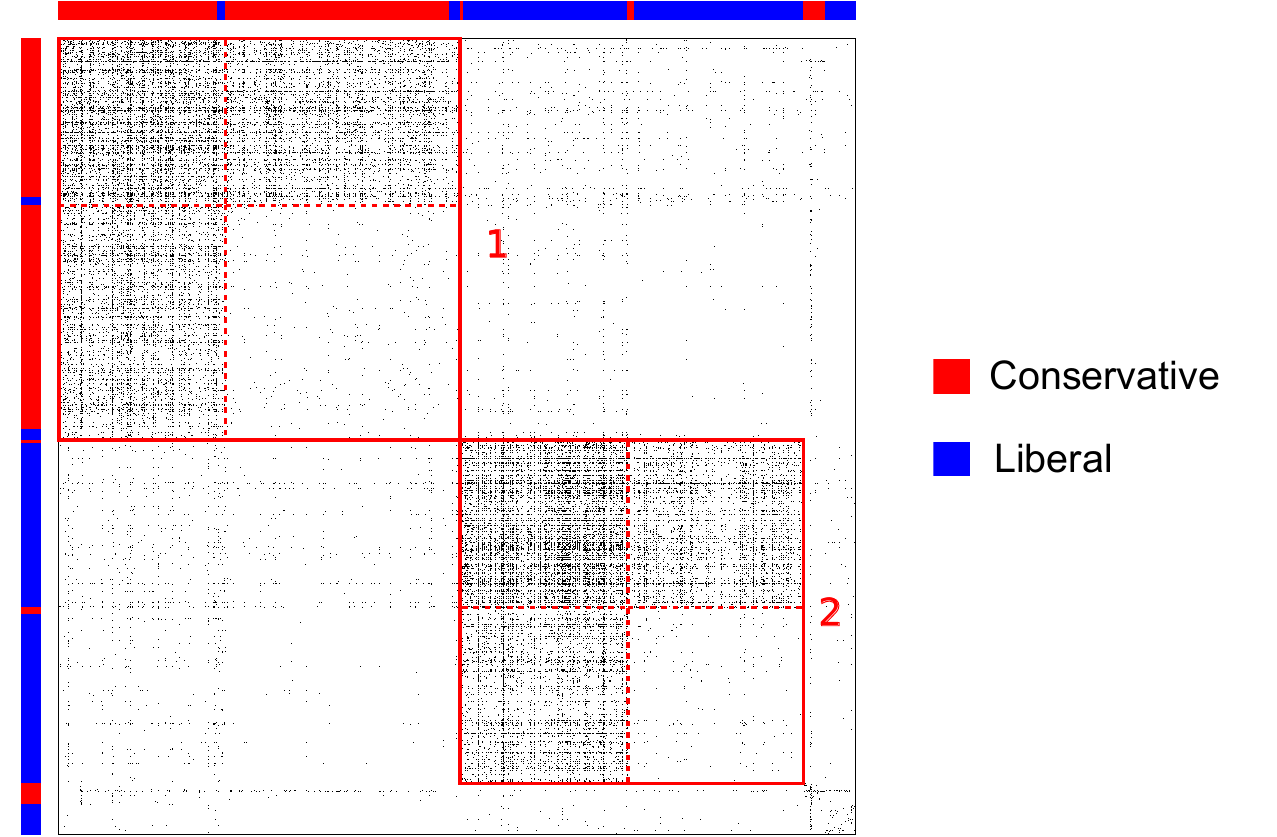}
	\caption{
		Significant core-periphery pairs in the political blog network.
    		The colour of the node indicates the political leaning of individual blogs. 
	}
	\label{fig:poliblog}
\end{figure}

For the blog network, KM--config identified two core-periphery pairs and 79 residual nodes (Fig.~\ref{fig:poliblog}).
A majority of the blogs leaning to the conservative and to the liberal belong to core-periphery pairs 1 and 2, respectively. 
These core-periphery pairs are similar to those identified by KM--ER \cite{Kojaku2017}.

\begin{figure}
	\centering
	\includegraphics[width=1.01\hsize]{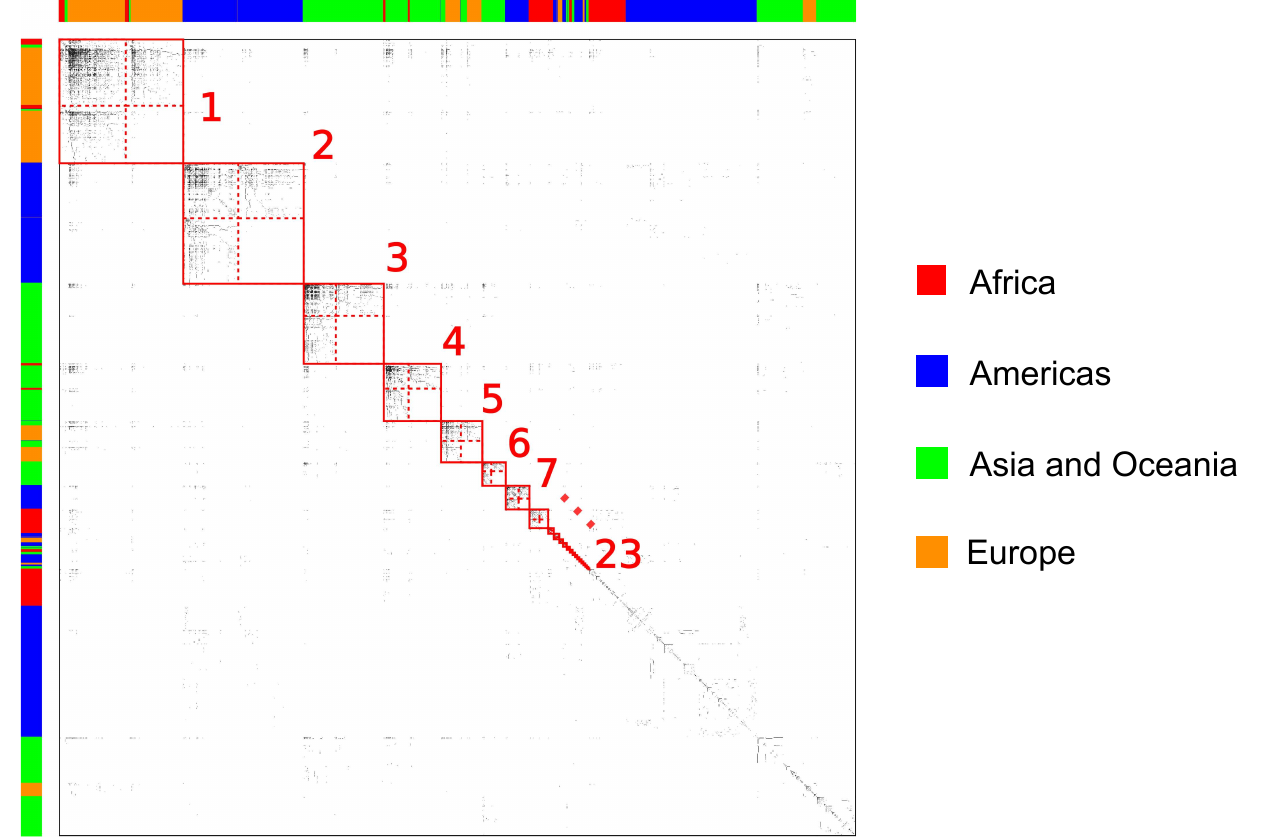}
	\caption{
		Significant core-periphery pairs in the airport network.
		The colour indicates the geographical region of the airports.
		Americas is the union of North, Central and South America.
	}
	\label{fig:airport}
\end{figure}

For the airport network, KM--config identified 23 significant core-periphery pairs and 983 residual nodes (Fig.~\ref{fig:airport}). 
Each core-periphery pair mainly consists of the airports in the same geographical region, which agrees with the previous results \cite{Guimera2005,Sales-Pardo2007,Kojaku2017}.
Our previous algorithm, KM--ER, detected ten core-periphery pairs, of which the three largest core-periphery pairs based in Europe, East Asia and the USA are similar to core-periphery pairs 1, 3 and 2 detected by KM--config, respectively \cite{Kojaku2017}.
Properties of core-periphery pairs 1--8 are shown in Table~\ref{ta:pairprofile}.
Among the representative airports (i.e., the airports having the largest degree in each core-periphery pair), 
some peripheral airports have a larger degree than core airports, e.g., MUC (Munich) in core-periphery pair 1, SVO (Moscow) in core-periphery pair 5 and NBO (Nairobi) in core-periphery pair 8, showing that hub nodes are not always classified as core nodes.

\subsection{Synthetic networks}
\begin{figure}
	\centering
	\includegraphics[width=0.5\hsize]{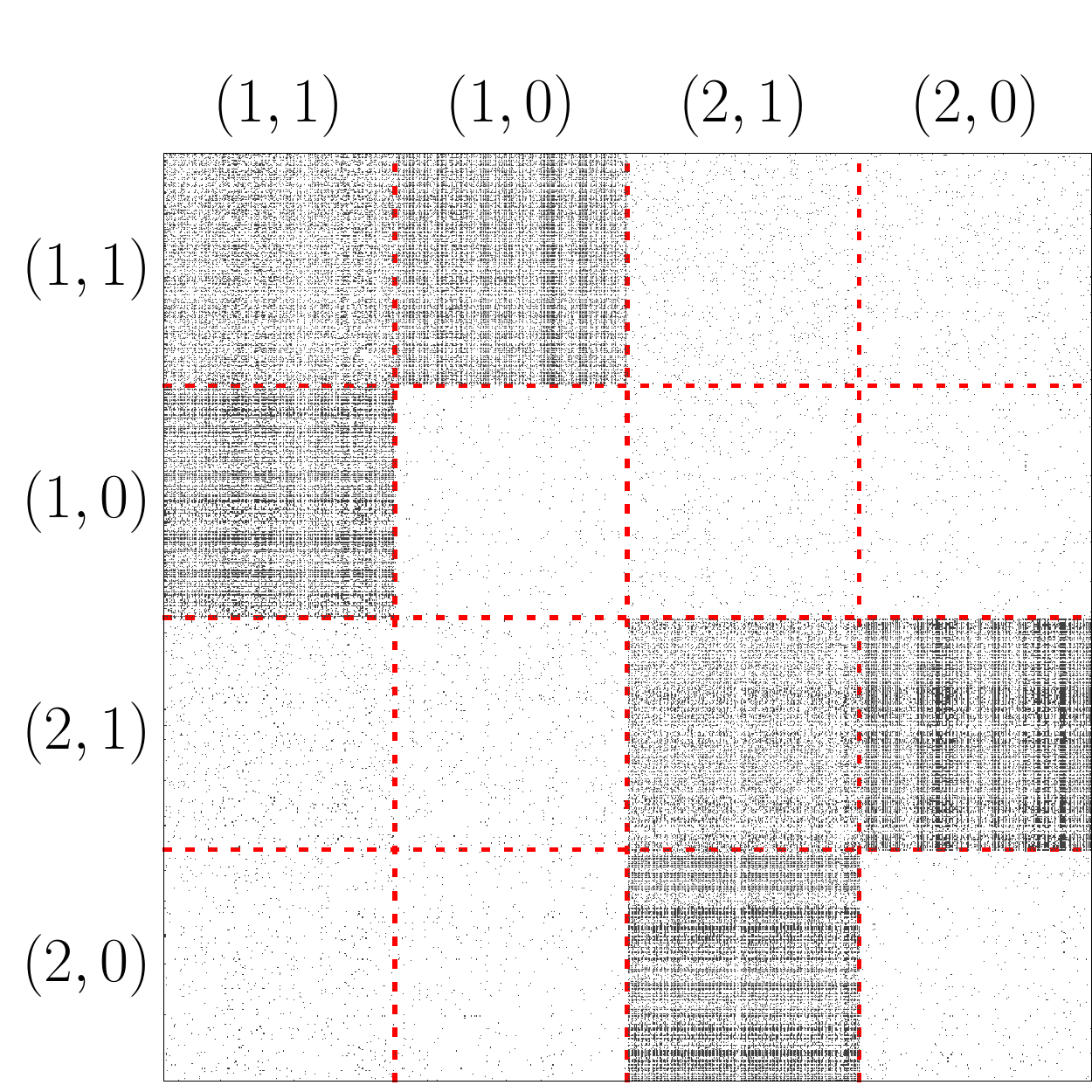}
	\caption{
		Adjacency matrix of a synthetic network generated with $\lambda=0.1$ and $\mu=0.7$.
		The dashed lines represent the boundaries between blocks.
		The label $(c, x)$ of each block is shown at the top and left of the adjacency matrix. 
	}
	\label{fig:synthe_adj}
\end{figure}

The results in the previous sections suggest that KM--config tends to detect core-periphery pairs without using the node's degree as a main criterion but produces node partitioning consistent with the concept of core-periphery structure based on edge density. To confirm this point further, in this section we test the algorithms on model networks with a planted core-periphery structure composed of two core-periphery pairs (Fig.~\ref{fig:synthe_adj}).

The discrepancy between the degree distribution of core nodes and that of peripheral nodes is controlled by a parameter $\mu \in \{0, 0.05, 0.1, \ldots, 0.5\}$.
The ``strength'' of the core-periphery structure is controlled by a parameter $\lambda \in \{0, 0.025, 0.05, \ldots, 1\}$.
The model assumes four blocks. Each block consists of 200 nodes and represents a core or periphery.
To generate networks, we use the degree-corrected SBM (dcSBM) \cite{Karrer2011}; it places edges such that 
each node $i$ has a prescribed expected degree $\overline d_i$, and each pair of blocks $u$ and $v$ has an expected number $\overline m_{uv}$ of edges.
We set $\overline d_i$ ($1 \leq i \leq N$) as follows.
For the core (i.e., blocks $(1,1)$ and $(2,1)$), we set $\overline d_i=50$ for a fraction $\mu$ of nodes and $\overline d_i=200$ for the remaining fraction $1-\mu$ of nodes.
For the periphery (i.e., blocks $(1,0)$ and $(2,0)$), we set $\overline d_i=50$ for a fraction $1-\mu$ of nodes and $\overline d_i=200$ for the remaining fraction $\mu$ of nodes.
The fraction $\mu$ tunes the amount of overlap between the degree distribution of core nodes and that of peripheral nodes.
The two distributions have no overlap if $\mu=0$ and perfectly overlap if $\mu=0.5$.
Then, we set $\overline m_{(c,x )(c',x')}$ ($1\leq c, c'\leq 2$ and $0\leq x, x'\leq 1$)  by
\begin{linenomath}
\begin{align}
	\label{eq:model_net}
	\overline m_{(c,x)(c',x')} = \lambda \overline m^{\text{\hspace{0.2em}rand}} _{(c,x)(c',x')} + (1-\lambda) \overline m^{\text{\hspace{0.2em}plant}}_{(c,x)(c',x')},
\end{align}
\end{linenomath}
where $\lambda$ is a mixing parameter, $\overline m^{\hspace{0.2em}\text{rand}} _{(c,x)(c',x')} = D_{(c,x)}D_{(c',x')}/2M$ is the expected number of edges between blocks $(c,x)$ and $(c',x')$ for the configuration model given $\overline d_1, \ldots, \overline d_N$.
The parameters $\{\overline m^{\text{\hspace{0.2em}plant}} _{(c,x),(c',x')}\}$ represent the number of intra-block and inter-block edges in an idealised core-periphery structure with nodes' degrees $\overline d_1, \ldots,\overline d_N$.
In other words, there are no edges between the different core-periphery pairs ($\overline m^{\text{\hspace{0.2em}plant}}_{(c,x),(c',x')}=0$ for $c\neq c'$), the
peripheral nodes are adjacent only to core nodes in the same core-periphery pair (i.e., $\overline m^{\text{\hspace{0.2em}plant}} _{(c,0),(c,1)} = D_{(c,0)}$ and $\overline m^{\text{\hspace{0.2em}plant}} _{(c,0),(c,0)} = 0$ for $c=1,2$), and the number of edges within each core is given by $\overline m^{\text{\hspace{0.2em}plant}} _{(c,1),(c,1)} = D_{(c,1)} - D_{(c,0)}$ for $c=1,2$.
Although the dcSBM with $\{ \overline m^{\text{\hspace{0.2em}plant}}_{(c,x),(c',x')} \}$ specifies a disconnected network, 
the dcSBM with $\{\overline m_{(c,x),(c',x')}\}$ given by Eq.~\eqref{eq:model_net} produces connected networks in general unless $\lambda = 0$.
We note that, if $\mu=0.5$, the dcSBM generates bipartite networks because the number of intra-core edges is zero, i.e., $\overline m^{\text{\hspace{0.2em}plant}} _{(c,1)(c,1)} = 0$ ($c=1,2$). 

We evaluate the performance of algorithms by the difference between the planted and detected core-periphery structures.
To quantify the difference, we use the variation of information (VI) \cite{Meila2007} given by 
\begin{linenomath}
\begin{align}
    \label{eq:vi}
    \text{VI}=
     -\sum_{(c,x)}\sum_{(\hat c, \hat x)} 
    R(c,x;\hat c, \hat x)
    \log 
    \frac{
        \left[R(c,x;\hat c, \hat x)\right]^2
    }{
        \left[\sum_{(\hat c',\hat x')}R(c,x;\hat c',\hat x')\right] \times
        \left[\sum_{(c',x')}R(c',x';\hat c,\hat x)\right]
    } 
    , 
\end{align}
\end{linenomath}
where $R(c,x;\hat c,\hat x)$ is the fraction of nodes having true label $(c,x)$ and inferred label $(\hat c,\hat x)$.
The VI value is the smallest (i.e., zero) if and only if the partitioning of nodes by the true labels and that by the inferred labels are identical. In the computation of the VI values, we regard the set of residual nodes as a block; technically, we set $(\hat c_i,\hat x_i) = (C+1,0)$ for the residual nodes. 
We generate 30 synthetic networks and average the VI values over the 30 generated networks.

\begin{figure}
	\centering
	\includegraphics[width=\hsize]{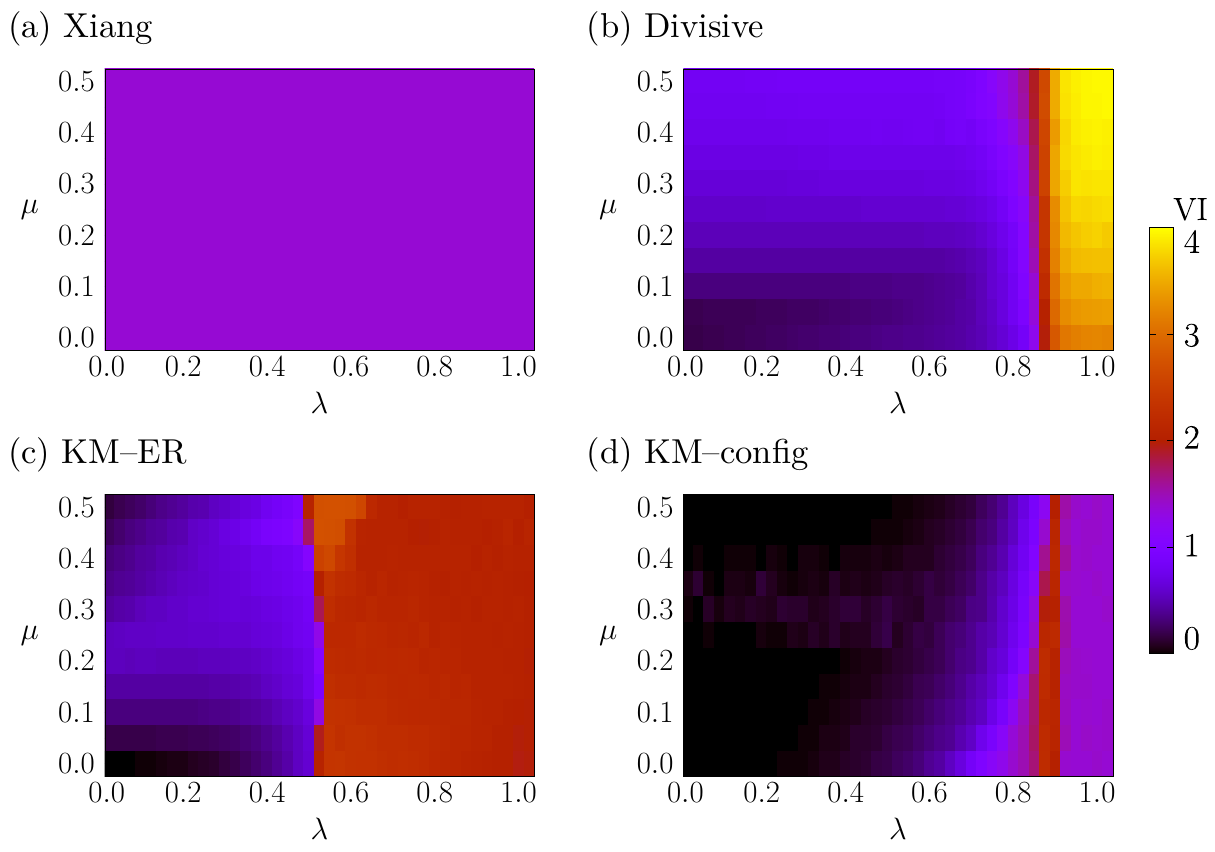}
	\caption{
		VI values between the planted and inferred core-periphery structure.
		(a) Xiang. 
		(b) Divisive.
		(c) KM--ER.
		(d) KM--config.
    	}
	\label{fig:synthe}
\end{figure}

The VI values for the Xiang, Divisive, KM--ER and KM--config algorithms  are shown in Fig.~\ref{fig:synthe}.
We do not show the results for the other three algorithms because they do not find multiple core-periphery pairs by definition.
When $\lambda$ is large, the VI values are large because the network is close to the configuration model and has weak core-periphery structure.
The VI values for the Xiang algorithm are large in the entire $\lambda$--$\mu$ parameter space (Fig.~\ref{fig:synthe}(a)).
This is because the Xiang algorithm did not find significant core-periphery pairs in all the generated networks (i.e., all nodes are residual nodes). 
The VI values for the Divisive and KM--ER algorithms are relatively large for most $\lambda$ values if some planted core nodes are non-hub nodes, i.e., $\mu \geq 0.15$ (Figs.~\ref{fig:synthe}(b) and \ref{fig:synthe}(c)). 
The VI values for the KM--config algorithm are the smallest in most of the $\lambda$--$\mu$ parameter space (Fig.~\ref{fig:synthe}(d)), including the case for bipartite-like structure ($\mu \geq 0.4$).
Therefore, KM--config but not the other algorithms is capable of detecting core-periphery structure even when a substantial fraction of core nodes are non-hubs and peripheral nodes are hubs.
\section{Discussion}
\label{sec:discussion}
We have studied core-periphery structure using the configuration model as the null model.
We have shown that discrete versions of a single core-periphery pair determined based on edge density, which many studies assume, can never be significant relative to the configuration model.
The core-periphery structure beyond what one expects for the configuration model must accompany other meso-scale network structure such as another core-periphery pair, communities and bipartite-like subnetworks coexisting in the given network.
This claim is in resonance with the studies \cite{IntVeld2014,Barucca2016,Kojaku2017} reporting the absence of core-periphery structure when the configuration model is used as the null model.
Then, we have presented a scalable algorithm to find core-periphery structure in networks and applied it to various networks. 

Our argument does not apply to continuous versions of core-periphery structure \cite{Borgatti2000,Boyd2010,Lee2014,Cucuringu2016,Rombach2017}, in which each node belongs to the core to a different extent.
A possible extension of our present algorithm (i.e., KM--config) to the case of continuous core-periphery structure is to replace the idealised core-periphery structure defined in Eq.~\eqref{eq:idealcp} with a continuous version of idealised core-periphery structure, such as those proposed in Refs.~\cite{Borgatti2000,Rombach2017}.
This line of investigation may reveal relationships between continuous versions of core-periphery structure, multiple core-periphery pairs and the configuration model.

Null models for networks do not have to be limited to the \ER random graph or the configuration model.
Other null models incorporate different properties of networks such as the weight of edges \cite{Mastrandrea2014}, the sign of edge weights \cite{Traag2009}, correlations \cite{MacMahon2015}, bipartiteness \cite{Barber2007} and spacial properties \cite{Expert2011}.
It is probably possible to incorporate such null models into our algorithm by modifying the null-model term of $\Qcp$ (i.e., the second term on the right-hand side of Eq.~\eqref{eq:q}).

Akin to the modularity, our quality function $\Qcp$ allows an interpretation in terms of random walks on networks.
With a core-periphery structure, random walkers are likely to move from any node to a core node within the same core-periphery pair in a discrete time step.
In a core-periphery structure on a small scale, the random walkers would reach the core in a small number of steps.
In contrast, they would need a large number of steps to reach the core on a large scale.
By regarding the number of steps as a resolution parameter, we may be able to identify core-periphery structure across different scales, as in the case of the
Markov stability, where modularity maximisation with different values of the time resolution parameter provides information about hierarchical organisation of communities in networks \cite{Delvenne2010,Mucha2010,Lambiotte2014,Masuda2017}.

Our quality function $\Qcp$ shares shortcomings with the modularity, such as the inability of finding small communities \cite{Fortunato2007} and of distinguishing random from non-random structure \cite{Guimera2004}.
Remedies for these problems include multi-resolution approaches \cite{Arenas2008} and statistical tests \cite{Lancichinetti2010,Zhang2014}.
Another approach is the statistical inference based on SBMs, which has been used for finding communities \cite{Karrer2011,Zhang2015,Barucca2016,Peixoto2017} and core-periphery structure \cite{Tunc2015,Zhang2015,Barucca2016}.
Investigation of core-periphery structure with SBMs may be a topic for future study. 

Finally, we have restricted ourselves to undirected and unweighted networks.
It is straightforward to incorporate the weight of edges by replacing $A_{ij}$ on the right-hand side of Eq.~\eqref{eq:q} by the weight of the edge between nodes $i$ and $j$.
In contrast, it is nontrivial to incorporate the direction of edges.
It seems that the direction of edges can be incorporated into $\Qcp$ by allowing an adjacency matrix to be asymmetric, as in the case of modularity \cite{Arenas2007,Leicht2008}.
However, for modularity, this extension elicits a problem \cite{Kim2010}, which may also hold true for $\Qcp$.

\appendix
\section{Finding compatible block structures}
\label{sec:compatibility}
We represent a block structure composed of $B$ blocks by a symmetric $B \times B$ matrix $\Ymat=(Y_{uv})$, 
where 
\begin{linenomath}
\begin{align}
	Y_{uv} = \left\{
		\begin{array}{cc}
		1 & \mbox{($m_{uv}  > \Exp\left[m_{uv} ^{\conf} \right]$)}, \\
		-1 & \mbox{($m_{uv}  < \Exp\left[m_{uv} ^{\conf} \right]$)}.
		\end{array}
		\right. \label{eq:Y}
\end{align}
\end{linenomath}
Variable $Y_{uv} = 1$ or $Y_{uv} = -1$ indicates that
blocks $u$ and $v$ are either densely (i.e., $m_{uv} > \Exp[m^{\conf} _{uv} ]$; filled blocks in Fig.~\ref{fig:blockmodel_bs2}) or sparsely  (i.e., $m_{uv} < \Exp[m^{\conf} _{uv} ]$; opened blocks in Fig.~\ref{fig:blockmodel_bs2}) interconnected.
Recall that we do not consider the case $m_{uv} = m^{\conf} _{uv}$ because it is unlikely in general.
To find compatible block structures, we generate all $2^{B(B+1)/2}$ symmetric binary matrices.
Then, for each binary matrix $\Ymat$, we inspect the compatibility of the block structure as follows.
Equation~\eqref{eq:balance} can be rewritten as 
\begin{linenomath}
\begin{align}
	\sum_{v=1}^B \Delta_{uv} = 0, \quad 1 \leq u \leq B, \label{eq:balance2p} 
\end{align}
\end{linenomath}
where
\begin{linenomath}
\begin{align}
	\Delta_{uv} = m_{uv} - \Exp[m_{uv} ^{\conf}]. \label{eq:delta}
\end{align}
\end{linenomath}
Equations~\eqref{eq:Y} and \eqref{eq:delta} imply 
\begin{linenomath}
\begin{align}
	Y_{uv}\Delta_{uv}>0,\quad 1\leq u, v \leq B. \label{eq:ydelta}
\end{align}
\end{linenomath}
Because $Y_{uv} \neq 0$, we rewrite Eq.~\eqref{eq:ydelta} as 
\begin{linenomath}
\begin{align}
	Y_{uv}\Delta_{uv} \geq 0,\quad 1\leq u, v \leq B,  \label{eq:ydelta1}\\\ 
	\Delta^2 _{uv} \geq \eta >0,\quad 1\leq u, v \leq B. \label{eq:ydelta2} \\
\end{align}
\end{linenomath}
We set $\eta = 1$ without loss of generality. 
In fact, if $\eta \neq 1$, consider a rescaled variable $\tilde \Delta_{uv}=\Delta_{uv} / \sqrt{\eta}$.
Dividing both sides of Eqs.~\eqref{eq:balance2p} and \eqref{eq:ydelta1} by $\sqrt{\eta}$ and of Eq.~\eqref{eq:ydelta2} by $\eta$ yields
\begin{linenomath}
\begin{align}
	\sum_{v=1}^B \tilde \Delta_{uv} = 0, \quad 1 \leq u \leq B, \\ 
	Y_{uv} \tilde \Delta_{uv} \geq 0,\quad 1\leq u, v \leq B, \\ 
	\tilde \Delta^2 _{uv} \geq 1,\quad 1\leq u, v \leq B,
\end{align}
\end{linenomath}
which are equivalent to Eqs.~\eqref{eq:balance2p}, \eqref{eq:ydelta1} and \eqref{eq:ydelta2} with $\eta=1$, respectively. 

We seek $\Delta_{uv}$ values ($1\leq u,v \leq B$) that simultaneously satisfy Eqs.~\eqref{eq:balance2p}, \eqref{eq:ydelta1} and \eqref{eq:ydelta2}.
To this end, we solve the following quadratic programming (QP) problem:
\begin{linenomath}
\begin{align}
	\min_{\substack{\Delta_{uv}; \\ 1\leq u,v\leq B}}  \sum_{u=1} \sum_{v=1} \Delta^2 _{uv}, \label{eq:objective}
\end{align}
\end{linenomath}
subject to Eqs.~\eqref{eq:balance2p}, \eqref{eq:ydelta1} and \eqref{eq:ydelta2}.
The constraints for the QP problem are equivalent to the conditions for compatible block structure. 
Therefore, the block structure is compatible if and only if the QP problem is feasible.
Note that we are not interested in the objective function's value, i.e., Eq.~\eqref{eq:objective}. 
To solve the QP problem and hence to check whether it has a feasible solution, we use a numerical solver \cite{gurobi}. 
If we find $\Delta_{uv}$ values satisfying all the constraints by solving the QP problem, then the block structure represented by $\Ymat = (Y_{uv})$ is compatible.
Otherwise, the block structure is incompatible.

\section{Network structure with four blocks}
\label{sec:all_patterns}
All possible network structures with four blocks that are compatible with Eq.~(\ref{eq:balance}) are shown in Fig.~\ref{fig:blockmodel_all}.

\section{Estimating statistical significance of core-periphery structure}
\label{sec:kde}
Let $S$ be the sum of the number of core-periphery pairs detected in the 500 randomised networks.  
Let $\hat q ^{(s)}$ and $\hat n ^{(s)}$ ($1\leq s\leq S$) be the quality and size of the $s$th core-periphery pair in the randomised networks, respectively. 
We use the Gaussian kernel density estimator \cite{Wand1993,Scott2012} to infer the joint probability distribution $P(\hat q, \hat n)$, which gives 
\begin{linenomath}
\begin{align}
    \label{eq:pqn}
    P(\hat q,\hat n) = \frac{1}{S} \sum_{s=1}^S 
        f\left( \frac{\hat q - \hat q^{(s)}}{\sigma_{\hat q}h}, \frac{\hat n - \hat n ^{(s)}}{\sigma_{\hat n}h} \right),    
\end{align}
\end{linenomath}
where $\sigma_{\hat q}$ and  $\sigma_{\hat n}$ are the standard deviations of $\{\hat  q^{(s)}\}$ and $\{\hat  n^{(s)}\}$  ($1 \leq s\leq S$), respectively.
Standard deviations $\sigma_{\hat q}$ and  $\sigma_{\hat n}$ are defined as 
\begin{linenomath}
\begin{align}
    \sigma_{\hat q} \equiv \sqrt{\frac{1}{S-1}\sum_{s=1}^S \left(\hat q^{(s)}-\langle \hat q \rangle\right)^2}, \\
    \sigma_{\hat n} \equiv \sqrt{\frac{1}{S-1}\sum_{s=1}^S \left(\hat n^{(s)}-\langle \hat n \rangle\right)^2},
\end{align}
\end{linenomath}
where $\langle \hat q \rangle$ and $\langle \hat n \rangle $ are the average values of $\{\hat q^{(s)}\}$ and  $\{\hat n^{(s)}\}$ ($1\leq  s \leq S$), respectively. 
In Eq.~\eqref{eq:pqn}, $f$ is the probability density function of the standard bivariate normal distribution given by 
\begin{linenomath}
\begin{align}
    \label{eq:bivariate}
    f(y_1,y_2) \equiv \frac{1}{2\pi \sqrt{1-\gamma ^2}} \exp\left( -\frac{ y_1 ^2  - 2 \gamma y_1 y_2 + y_2 ^2 }{2\left(1-\gamma^2 \right)}  \right), 
\end{align}
\end{linenomath}
where $\gamma$ is the Pearson correlation coefficient between $\{ \hat q^{(s)} \}$ and $\{ \hat n ^{(s)} \}$ ($1 \leq s \leq S$), i.e.,  
\begin{linenomath}
\begin{align}
    \label{eq:pearson}
    \gamma \equiv \frac{\displaystyle \frac{1}{S}\sum_{s=1} ^S \left(\hat q^{(s)}-\langle \hat  q \rangle \right)\left(\hat n^{(s)}-\langle \hat  n \rangle \right)}{ \displaystyle \sigma_{\hat q} \sigma_{\hat n}}.    
\end{align}
\end{linenomath}
In Eq.~\eqref{eq:pqn}, $h$ is a parameter specifying the width of the Gaussian kernel density estimator. 
We use the Scott's rule of thumb \cite{Scott2012}, which gives $h=S^{-1/6}$.
 
The probability that the core-periphery pair of size $n$ has a quality value greater than or equal to $q$ in randomised networks is computed as
\begin{linenomath}
\begin{align}
    \label{eq:pval0}
    P(\hat q \geq q \given n) &=  \dfrac{\displaystyle \int ^{\infty} _{q} P( z , n ) {\rm d}z}{\displaystyle \int ^{\infty} _{-\infty} P(z, n)  {\rm d}z}
    =  \dfrac{\displaystyle \sum\limits_{s=1}^{S} \int ^{\infty} _{q} f\left( \frac{z - \hat q^{(s)}}{\sigma_{\hat q}h}, \frac{ n - \hat n ^{(s)}}{\sigma_{\hat n}h} \right) {\rm d}z}{\displaystyle \sum\limits_{s=1}^{S} \int ^{\infty} _{-\infty} f\left( \frac{z - \hat q^{(s)}}{\sigma_{\hat q}h}, \frac{n - \hat n ^{(s)}}{\sigma_{\hat n}h} \right) {\rm d}z}.    
\end{align}
\end{linenomath}
Equation~\eqref{eq:bivariate} leads to 
\begin{linenomath}
\begin{align}
    \label{eq:bivariatecum}
    \int_{-\infty} ^{y_1} f\left( z , y_2 \right){\rm d}z = \frac{1}{\sqrt{2\pi }}\exp\left(-\frac{y_2 ^2}{2}\right)\Phi\left( \frac{ y_1-\gamma y_2 }{\sqrt{1-\gamma^2}} \right),
\end{align}
\end{linenomath}
where $\Phi\left( y \right) = (2\pi)^{-1/2}\int^{y} _{-\infty} \exp(-z^2 /2){\rm d}z$ is the cumulative distribution function of the standard normal distribution.
Substitution of Eq.~\eqref{eq:bivariatecum} into Eq.~\eqref{eq:pval0} yields 
\begin{linenomath}
\begin{align}
    \label{eq:pval}
    P(\hat q \geq q \given n) &= 1 - \dfrac{
                    \displaystyle \sum\limits_{s=1}^{S}{ \exp\left( -\frac{\left(n - \hat n ^{(s)} \right)^2}{2\sigma_{\hat n}^2h^2}\right)}
                    \Phi\left( \frac{\sigma_{\hat n}\left(q -\hat q^{(s)}\right)  - \gamma \sigma_{\hat q}\left(n-\hat n^{(s)}\right)}{\sigma_{\hat n}\sigma_{\hat q}h\sqrt{1-\gamma^2}}\right)
                    }{
                    \displaystyle \sum\limits_{s=1}^{S}{\exp\left( -\frac{\left(n -  \hat n^{(s)} \right)^2}{2\sigma_{\hat n}^2h^2}\right)}
                    }.    
\end{align}
\end{linenomath}

\section{Results of other algorithms}
\label{sec:results_of_other_algorithms}
\begin{figure}
	\centering
	\includegraphics[width=1\hsize]{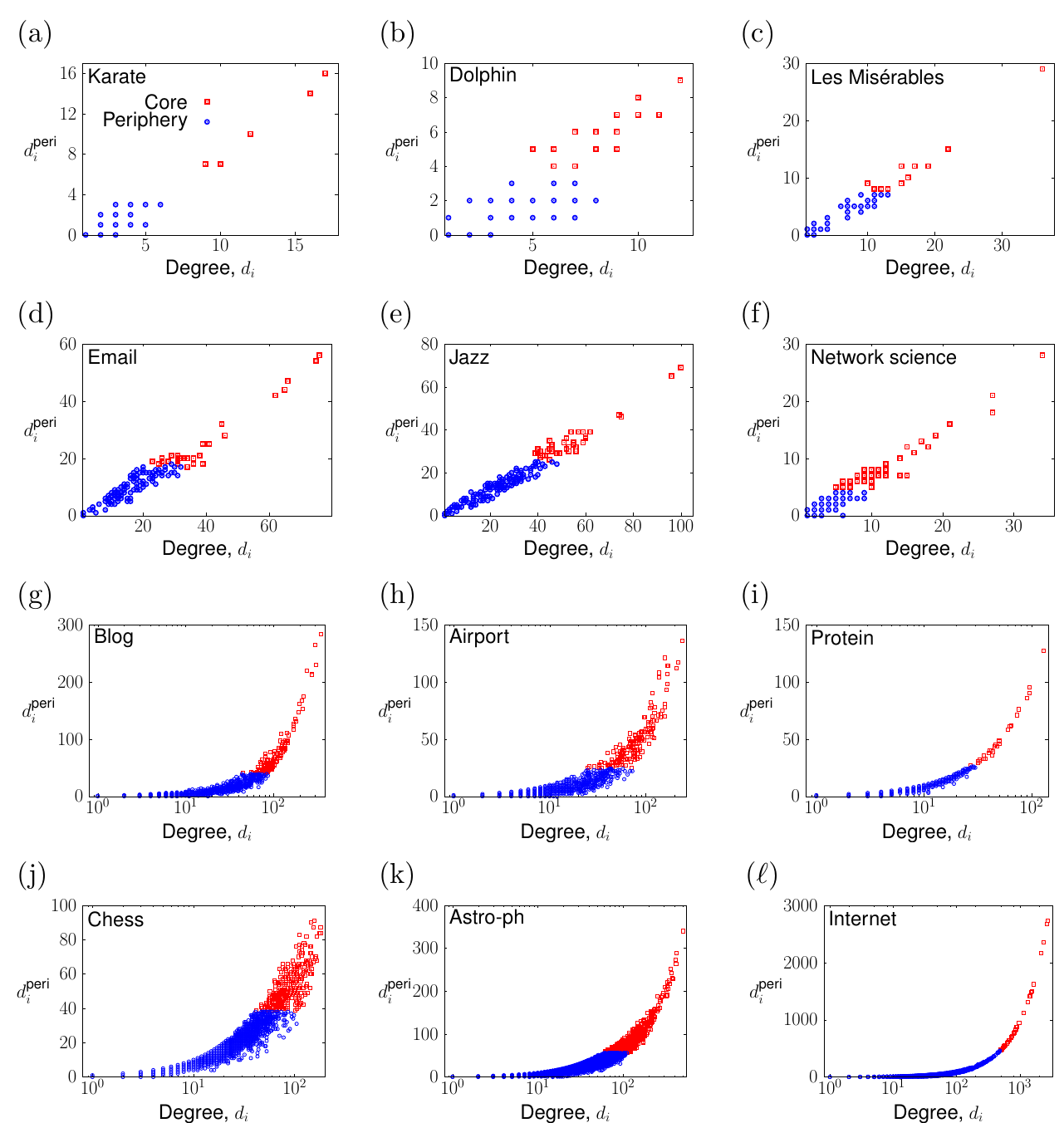}
	\caption{
		Relationships between $d_i$ and $d^{\text{peri}} _i$ for the empirical networks when the core-periphery structure is determined by the BE algorithm.
		The squares and circles indicate core nodes and peripheral nodes, respectively.
	}
	\label{fig:neighbours_be}
\end{figure}

\begin{figure}
	\centering
	\includegraphics[width=1\hsize]{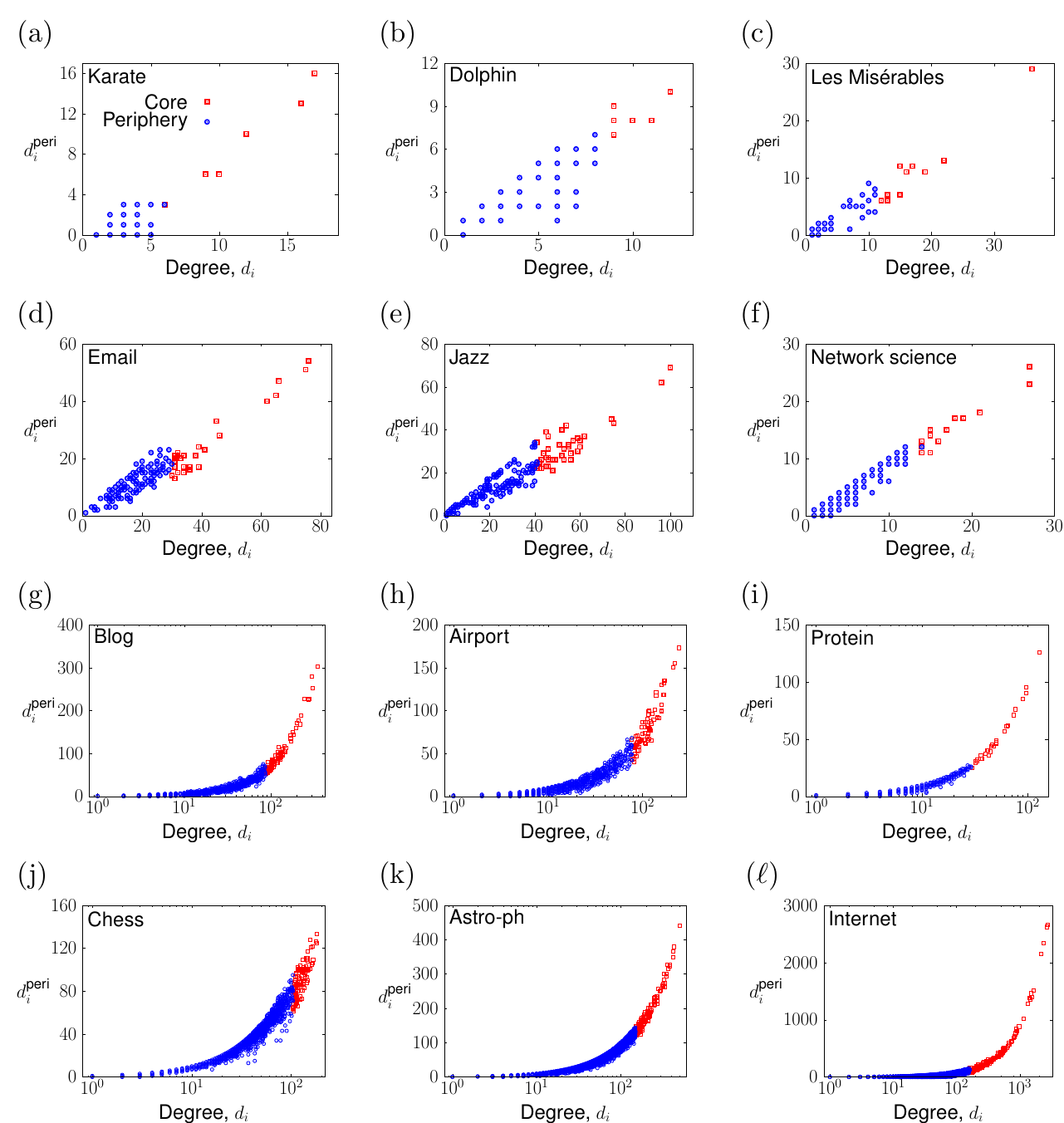}
	\caption{
		Relationships between $d_i$ and $d^{\text{peri}} _i$ for the empirical networks when the core-periphery structure is determined by the MINRES algorithm.
	}
	\label{fig:neighbours_minres}
\end{figure}

\begin{figure}
	\centering
	\includegraphics[width=1\hsize]{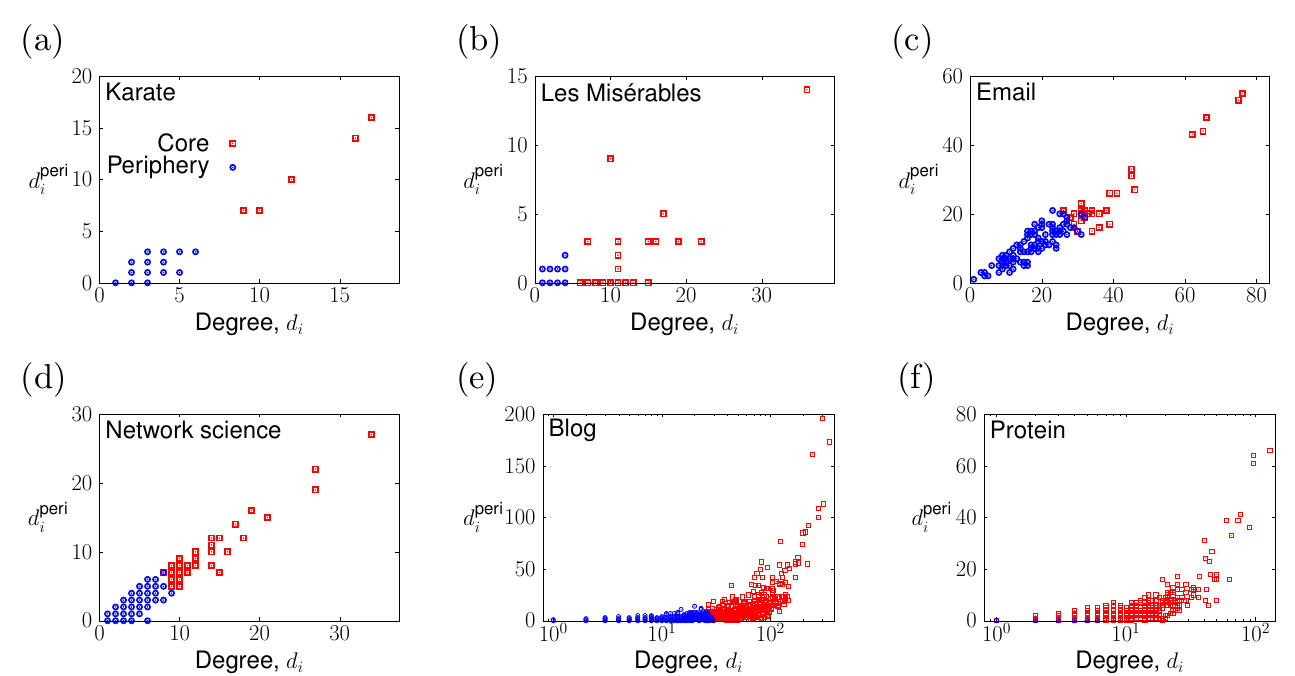}
	\caption{
		Relationships between $d_i$ and $d^{\text{peri}} _i$ for the empirical networks when the core-periphery structure is determined by the SBM algorithm.
		We do not show the results for the other empirical networks, for which the SBM algorithm does not find significant core-periphery pairs.
	}
	\label{fig:neighbours_zm}
\end{figure}

\begin{figure}
	\centering
	\includegraphics[width=0.66\hsize]{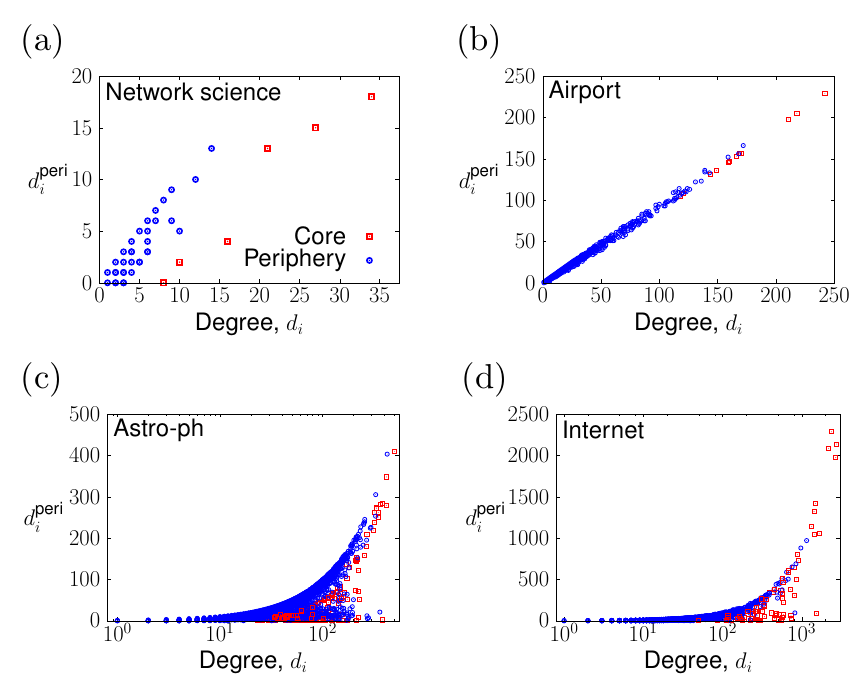}
	\caption{
		Relationships between $d_i$ and $d^{\text{peri}} _i$ for the empirical networks when the core-periphery structure is determined by the Xiang algorithm.
		We do not show the results for the other empirical networks, for which the Xiang algorithm does not find significant core-periphery pairs.
	}
	\label{fig:neighbours_xian}
\end{figure}

\begin{figure}
	\centering
	\includegraphics[width=1\hsize]{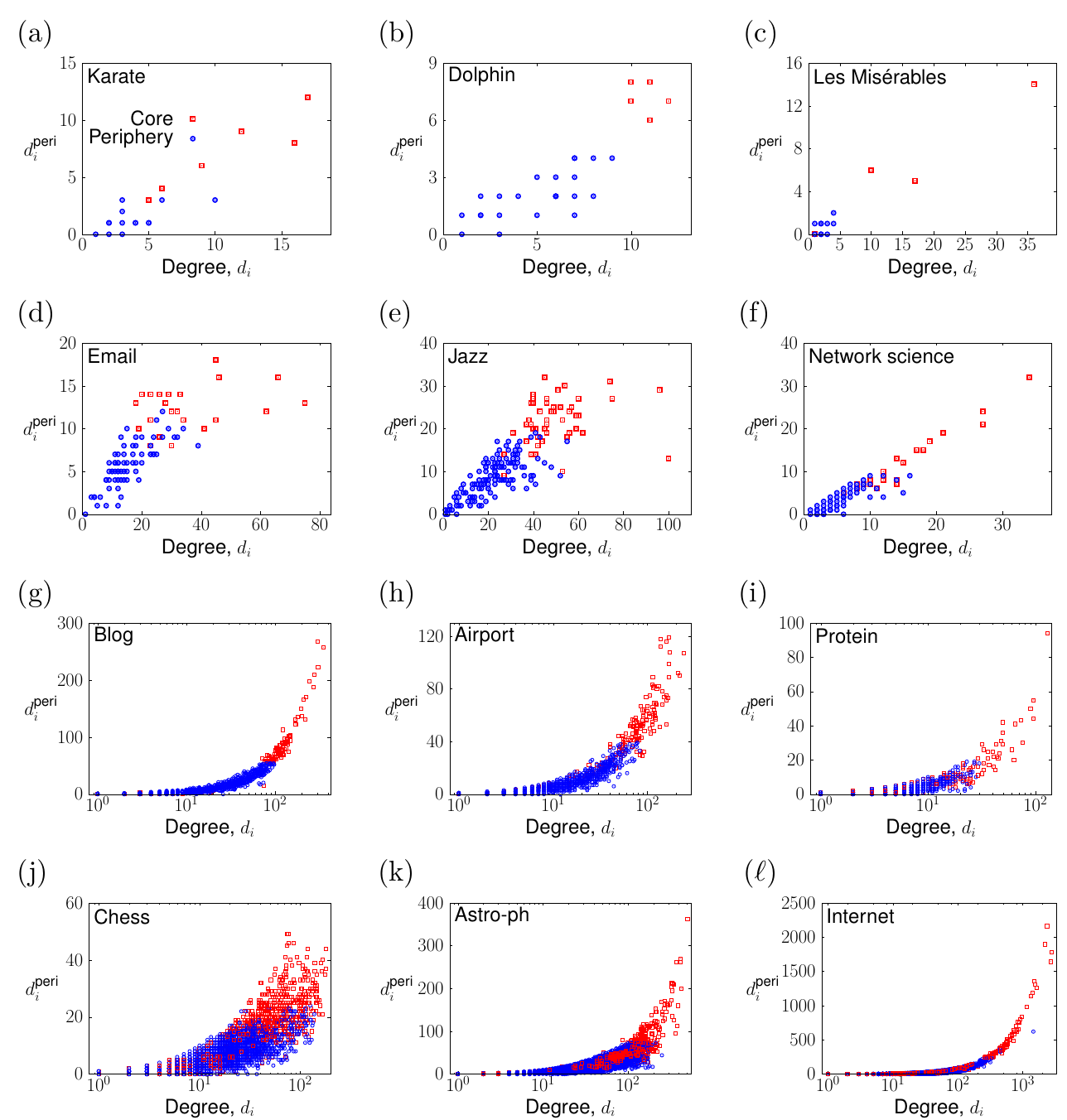}
	\caption{
		Relationships between $d_i$ and $d^{\text{peri}} _i$ for the empirical networks when the core-periphery structure is determined by the Divisive algorithm.
	}
	\label{fig:neighbours_dv}
\end{figure}

\begin{figure}
	\centering
	\includegraphics[width=1\hsize]{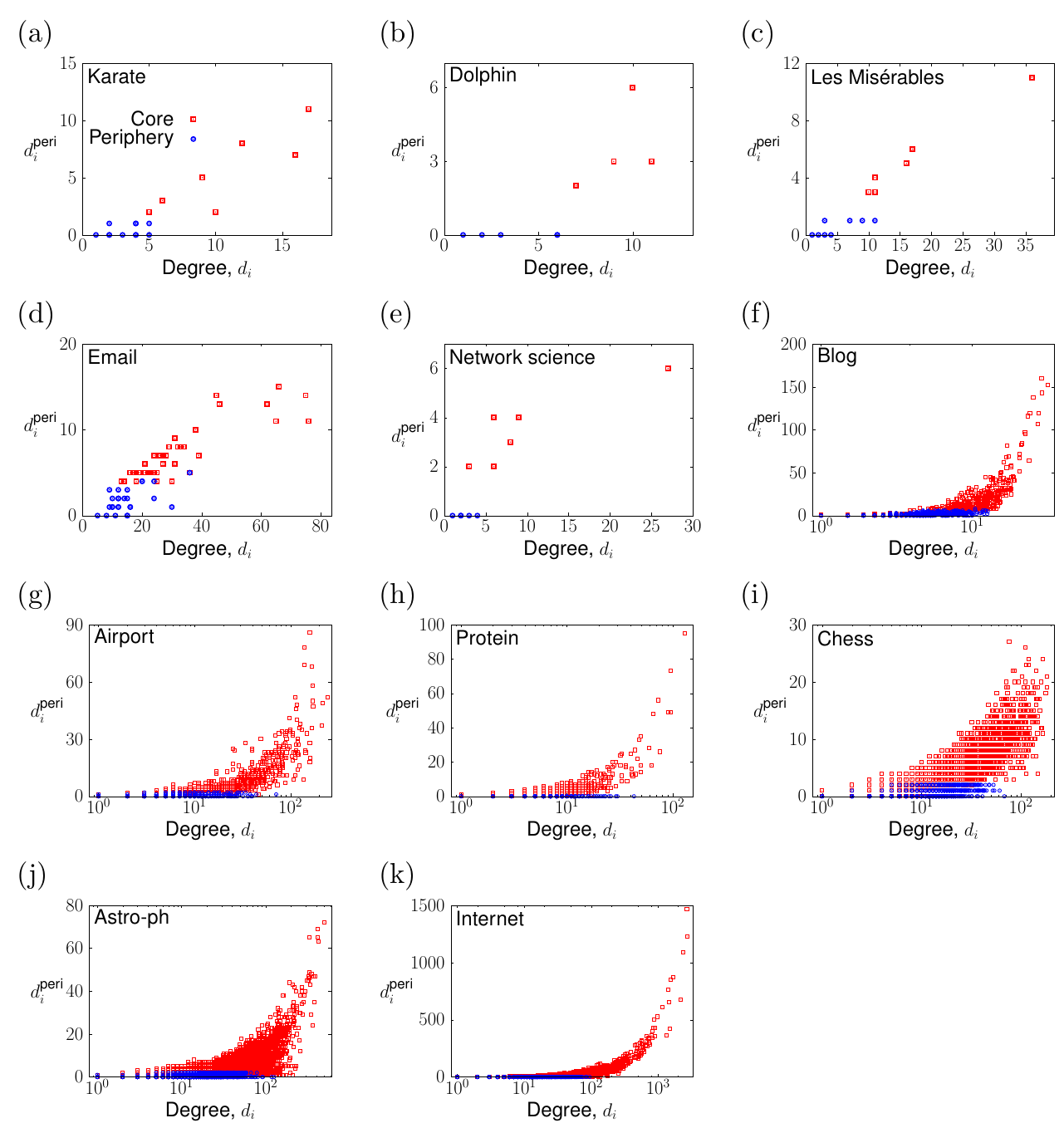}
	\caption{
		Relationships between $d_i$ and $d^{\text{peri}} _i$ for the empirical networks when the core-periphery structure is determined by KM--ER.
		We do not show the result for the jazz network, for which KM--ER does not find significant core-periphery pairs.
	}
	\label{fig:neighbours_kmer}
\end{figure}

\begin{figure}
	\centering
	\includegraphics[width=1\hsize]{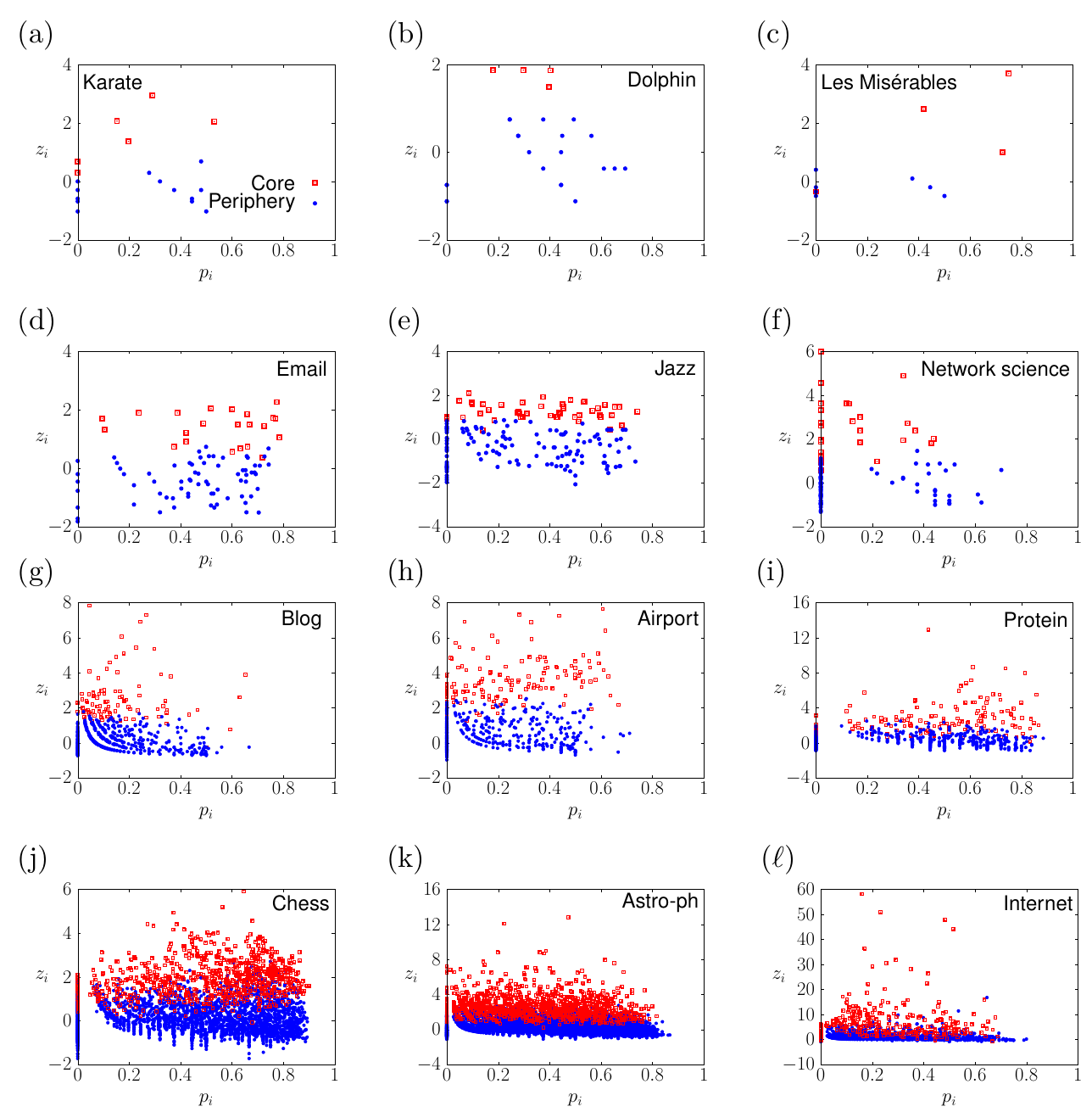}
	\caption{
		Cartographic analysis of the empirical networks.
		The core and peripheral nodes are detected by the Divisive algorithm.
	}
	\label{fig:cartography_dv}
\end{figure}

Figures \ref{fig:neighbours_be}--\ref{fig:neighbours_kmer} plot $d^{\text{peri}} _i$ against degree $d_i$ when the core-periphery structure is determined by the BE, MINRES, SBM, Xiang, Divisive and KM--ER algorithms. 
Figure \ref{fig:cartography_dv} shows the cartographic representation of the 12 empirical networks when the core-periphery structure is determined by the Divisive algorithm.

\bibliographystyle{apsrev4-1}
%

\clearpage
\clearpage
\clearpage

\clearpage
	
\clearpage

\clearpage
%
\clearpage
%
\clearpage
\clearpage

\clearpage
\clearpage
\clearpage
\clearpage
\clearpage
\clearpage
\clearpage
\clearpage
		
\clearpage
	
\clearpage

\clearpage
	
\clearpage
\clearpage
\clearpage

\clearpage

\clearpage

\begin{table}
\centering
\caption{
	Empirical networks used in the present paper: 
	the karate club network \cite{Zachary1977}, 
	dolphin social network \cite{Lusseau2003}, 
	network of characters in Les Mis\'{e}rables \cite{Knuth1993},
	Enron email network \cite{Klimt2004},
	network of jazz musicians \cite{Gleiser2003},
	co-authorship network in network science \cite{Newman2006}, 
	political blog network \cite{Adamic2005}, 
	worldwide airport network \cite{Openflight.org,ToreOpsahl},
	protein-protein interaction network \cite{Rual2005},
	network of chess players \cite{KONECT},
	co-authorship network in the arXiv astro-ph section \cite{Leskovec2007} and 
	the Internet at the level of AS \cite{KONECT}. 
	We exclude isolated nodes and self-loops from the networks.
	We count the multi-edges between a pair of nodes as a single edge. 
}
\label{ta:net_stat}
\scalebox{0.9}{
\begin{tabular}{l|ccccc}
\multirow{2}{*}{Network} & \multirow{2}{4em}{\centering $N$} & \multirow{2}{4em}{\centering $M$} & \multirow{2}{7em}{\centering Assortativity} & \multicolumn{2}{c}{Degree} \\ \cline{5-6} 
& & &  & \multirow{1}{5em}{\centering Average} &  \multirow{1}{5em}{\centering Maximum} 
\\ \hline \hline
Karate  \cite{Zachary1977} 				& 34 		& 78 		     & $-0.475$ & 4.59 	& 17 	   \\
Dolphin  \cite{Lusseau2003}				& 62 		& 159 		     & $-0.044$ & 5.13 	& 12 	   \\
Les Mis\'{e}rables  \cite{Knuth1993} 			& 77 		& 254 		     & $-0.165$ & 6.60 	& 36 	   \\
Email \cite{Klimt2004} 					& 151 		& 1${,}$527 	     & $-0.059$ & 20.23 & 74 	   \\
Jazz \cite{Gleiser2003} 				& 198 		& $2{,}742$ 	     & $0.020$  & 27.70 & 100 	   \\
Network science (co-authorship) \cite{Newman2006} 	& 379 		& 914 		     & $-0.082$ & 4.82 	& 34 	   \\
Blog \cite{Adamic2005} 					& 1${,}$222 	& 16${,}$714 	     & $-0.221$ & 27.36 & 351 	   \\
Airport \cite{Openflight.org,ToreOpsahl} 		& 2${,}$939 	& 15${,}$677 	     & $0.051$  & 10.67 & 242 	   \\
Protein \cite{Rual2005} 				& 3${,}$023 	& 6${,}$149 	     & $-0.126$ & 4.07 	& 129 	   \\
Chess \cite{KONECT} 					& $7{,}115$ 	& $55{,}779$ 	     & $0.371$  & 15.68 & 181 	   \\
Astro-ph (co-authorship) \cite{Leskovec2007} 		& 18${,}$771 	& $198${,}$050$      & $0.205$  & 21.10 & 504 	   \\
Internet \cite{KONECT} 					& 34${,}$761 	& 107${,}$720 	     & $-0.215$ & 6.20 	& 2${,}$760\\ \hline 
\end{tabular}
}
\end{table}
\clearpage

\begin{table}
\centering
\caption{
AUCs of the ROC curves shown in Fig.~\ref{fig:roc}.  
The asterisk indicates that the algorithm does not detect significant core-periphery pairs.
}
\label{ta:auc}
\begin{tabular}{l|cccccccc}
\multirow{2}{8em}{Network} 		& \multirow{2}{4em}{\centering BE}	& \multirow{2}{4em}{\centering MINRES}& \multirow{2}{4em}{\centering SBM}	& \multirow{2}{4em}{\centering Xiang}& \multirow{2}{4em}{\centering Divisive}& \multirow{2}{4em}{\centering KM--ER} & \multirow{2}{4em}{\centering KM--config} \\ \\ \hline \hline
Karate 			& 1.000 & 1.000 & 1.000 & * 	& 0.957 & 0.984 & 0.938\\
Dolphin 		& 0.953 & 1.000 & * 	& * 	& 1.000 & 1.000 & 0.859\\
Les Mis\'{e}rables 	& 0.982 & 1.000 & 1.000 & * 	& 0.886 & 0.955 & 0.610\\
Email 			& 0.978 & 0.999 & 0.990 & * 	& 0.910 & 0.893 & 0.670\\
Jazz 			& 0.989 & 1.000 & * 	& * 	& 0.953 & * 	& 0.717\\
Network science 	& 0.979 & 1.000 & 0.998 & 0.958 & 0.961 & 0.990 & 0.664\\
Blog 			& 0.995 & 1.000 & 0.999 & * 	& 0.981 & 0.932 & 0.718\\
Airport 		& 0.996 & 1.000 & * 	& 0.999 & 0.972 & 0.885 & 0.793\\
Protein 		& 1.000 & 1.000 & 0.824 & * 	& 0.936 & 0.810 & 0.717\\
Chess 			& 0.997 & 1.000 & * 	& * 	& 0.928 & 0.860 & 0.737\\
Astro-ph 		& 0.997 & 1.000 & * 	& 0.883 & 0.943 & 0.888 & 0.834\\
Internet 		& 1.000 & 1.000 & * 	& 0.999 & 0.972 & 0.905 & 0.483\\ \hline
\end{tabular}
\end{table}
\clearpage

\begin{table}
\centering
\caption{
	Modularity for communities determined by the Louvain algorithm and that for the core-periphery pairs determined by KM--config.
}
\label{ta:vscommunity}
\begin{tabular}{p{8em}|cc}
Network			& \multirow{1}{7em}{\centering Louvain}	& \multirow{1}{7em}{\centering KM--config} \\ \hline
Karate 			& 0.416 	& 0.417 \\  
Dolphin 		& 0.520 	& 0.518 \\
Les Mis\'{e}rables 	& 0.535 	& 0.542 \\
Email 			& 0.420 	& 0.419 \\
Jazz 			& 0.445 	& 0.445 \\
Network science 	& 0.815 	& 0.741 \\
Blog 			& 0.426 	& 0.426 \\
Airport 		& 0.642 	& 0.615 \\
Protein 		& 0.626 	& 0.483  \\
Chess 			& 0.505 	& 0.508 \\
Astro-ph 		& 0.574 	& 0.555 \\
Internet 		& 0.521 	& 0.459 \\ \hline
\end{tabular}
\end{table}
\clearpage

\begin{table}
\centering
\caption{
Property of the eight largest significant core-periphery pairs in the airport network.
The representative airports of each core-periphery pair are defined as the four core and four peripheral airports having the largest degree.
The territory is defined as the country where the airport is located.
If the airport is located in a sovereign state, we instead show the name of the state.
IATA is a three-letter code of an airport assigned by the International Air Transport Association.
}
\label{ta:pairprofile}
\scalebox{0.72}{
\begin{tabular}{cccllp{8em}lcllp{8em}lc}
\multirow{2}{4em}{\centering Pair} & 
\multicolumn{2}{c}{Number of airports} & 
&
\multicolumn{4}{c}{Representative core airport} & 
& 
\multicolumn{4}{c}{Representative peripheral airport}\\ \cline{2-3} \cline{5-8}\cline{10-13}
& 
Core & 
Periphery & 
& 
IATA & City & Territory & Degree & 
& 
IATA & City & Territory & Degree \\ \hline \hline 
\multirow{4}{*}{1} & \multirow{4}{*}{245} & \multirow{4}{*}{212} &   & FRA & Frankfurt & Germany 	& 242 &	  & MUC & Munich & Germany& 149 \\
		 &  &  &   & CDG & Paris & France 	& 218 &	   & OSL & Oslo & Norway& 91  \\
		 &  &  &   & AMS & Amsterdam & Netherlands 	& 211&	   & BUD & Budapest & Hungary& 77 \\
		 &  &  &   & LGW & London & UK 	& 172&	   & LPL & Liverpool & UK& 66 \\ \hline	
\multirow{4}{*}{2} & \multirow{4}{*}{203} & \multirow{4}{*}{242} &   & ATL & Atlanta & USA 	& 168 &	   & MEX & Mexico City & Mexico& 81 \\
		 &  &  &   & JFK & New York & USA 	& 144 &	   & LGA & New York & USA& 44 \\
		 &  &  &   & LAS & Las Vegas & USA 	& 139 &	   & CCS & Caracas & Venezuela& 41 \\
		 &  &  &   & YYZ & Toronto & Canada 	& 119 &	   & SXM & Philipsburg & Netherlands Antilles & 35 \\ \hline	
\multirow{4}{*}{3} & \multirow{4}{*}{118} & \multirow{4}{*}{177} &   & PEK & Beijing & China 	& 170 & & HGH & Hangzhou & China& 53 \\
		 &  &  &   & BKK & Bangkok & Thailand 	& 136 &	   & KIX & Osaka & Japan& 47 \\
		 &  &  &   & PVG & Shanghai & China 	& 126 &	   & NGO & Nagoya & Japan& 32 \\
		 &  &  &   & ICN & Seoul & South Korea 	& 121 &	   & WNZ & Wenzhou & China& 32 \\ \hline	
\multirow{4}{*}{4} & \multirow{4}{*}{92} & \multirow{4}{*}{120} &   & DXB & Dubai & UAE 	&  166	 && BAH & Bahrain & Bahrain& 52 \\
		 &  &  &   & JED & Jeddah & Saudi Arabia 	& 99	   && MRU &  Port Louis & Mauritius& 29 \\
		 &  &  &   & DEL & Delhi & India 	& 93	&   & MLE & Mal\'{e} & Maldives& 27 \\
		 &  &  &   & DOH & Doha & Qatar 	& 92	&   & KBL & Kabul & Afghanistan& 23 \\ \hline	
\multirow{4}{*}{5} & \multirow{4}{*}{73} & \multirow{4}{*}{79} &   & DME & Moscow & Russia 	& 159	&& SVO & Moscow & Russia& 118 \\
		 &  &  &   & LED & St. Petersburg & Russia 	& 92	   && DYU & Dushanbe & Tajikistan& 26 \\
		 &  &  &   & KBP & Kiev & Ukraine 	& 76	   && ODS & Odessa & Ukraine& 15 \\
		 &  &  &   & TLV & Tel-aviv & Israel 	& 69	   && YKS & Yakutsk & Russia& 14 \\ \hline	
\multirow{4}{*}{6} & \multirow{4}{*}{33} & \multirow{4}{*}{54} &   & SYD & Sydney & Australia 	& 77	&   & DRW & Darwin & Australia& 20 \\
		 &  &  &   & MEL & Melbourne & Australia 	& 56	&   & TSV & Townsville & Australia& 11 \\
		 &  &  &   & BNE & Brisbane & Australia 	& 51	 &  & NOU & {Noum\'{e}a} & New Caledonia & 10 \\
		 &  &  &   & AKL & Auckland & New Zealand 	& 38	 &  & WLG & Wellington & New Zealand& 6 \\ \hline	
\multirow{4}{*}{7} & \multirow{4}{*}{47} & \multirow{4}{*}{40} &   & GRU & S\~{a}o Paulo & Brazil 	& 83	&   & BEL & Bel\'{e}m & Brazil& 17 \\
		 &  &  &   & GIG & Rio De Janeiro & Brazil 	& 46	&   & CGR & Campo Grande & Brazil& 16 \\
		 &  &  &   & BSB & Bras\'{i}lia & Brazil 	& 46 &	   & MVD & Montevideo & Uruguay& 13 \\
		 &  &  &   & CNF & Belo Horizonte & Brazil 	& 33 &	   & NAT & Natal & Brazil& 12 \\ \hline	
\multirow{4}{*}{8} & \multirow{4}{*}{37} & \multirow{4}{*}{32} &   & JNB & Johannesburg & South Africa 	& 70	&   & NBO & Nairobi & Kenya& 68 \\
		 &  &  &   & ADD & Addis Ababa & Ethiopia 	& 58	&   & WDH & Windhoek & Namibia& 9 \\
		 &  &  &   & LAD & Luanda & Angola 	& 30	&   & PNR & Pointe-noire & Congo (Republic) & 8 \\
		 &  &  &   & DAR & Dar Es Salaam & Tanzania 	& 23	 &  & NDJ & N'djamena & Chad& 8 \\ \hline	
\end{tabular}
}
\end{table}
\end{document}